\documentclass[12pt]{article}
\usepackage{epsfig}
\usepackage{psfrag}
\usepackage{latexsym}
\usepackage{indentfirst}
\usepackage{fancyhdr}
\usepackage{amssymb}
\usepackage{amsfonts}
\usepackage{cite}
\usepackage{bbold}
\usepackage{color}
\usepackage[footnotesize]{caption2}
\usepackage{graphicx}
\usepackage[center,footnotesize,hang]{subfigure}
\usepackage{url}
\newcommand{\be}{\beta}

\newcommand{\beq}{\begin{equation}}
\newcommand{\eeq}{\end{equation}}
\newcommand{\bac}{\beq\begin{array}}
\newcommand{\eac}{\end{array}\eeq}
\newcommand{\ba}{\begin{array}}
\newcommand{\ea}{\end{array}}
\newcommand{\bea}{\begin{eqnarray}}
\newcommand{\eea}{\end{eqnarray}}
\newcommand{\beaa}{\begin{eqnarray*}}
\newcommand{\eeaa}{\end{eqnarray*}}
\def\beq{\begin{equation}}
\def\eeq{\end{equation}}
\def\bea{\begin{eqnarray}}
\def\eea{\end{eqnarray}}
\def\bet{\begin{tabular}}
\def\eet{\end{tabular}}
\def\bes{\begin{subequations}\bea}
\def\ees{\eea\end{subequations}}

\def\be{\begin{equation}}
\def\ee{\end{equation}}
\def\bc{\begin{center}}
\def\ec{\end{center}}
\def\bea{\begin{eqnarray}}
\def\eea{\end{eqnarray}}

\catcode`@=11
\def\marginnote#1{}
\newcount\hour
\newcount\minute
\newtoks\amorpm
\hour=\time\divide\hour by60
\minute=\time{\multiply\hour by60 \global\advance\minute by-\hour}
\edef\standardtime{{\ifnum\hour<12 \global\amorpm={am}%
        \else\global\amorpm={pm}\advance\hour by-12 \fi
        \ifnum\hour=0 \hour=12 \fi
        \number\hour:\ifnum\minute<10 0\fi\number\minute\the\amorpm}}
\edef\militarytime{\number\hour:\ifnum\minute<10 0\fi\number\minute}
\def\draftlabel#1{{\@bsphack\if@filesw {\let\thepage\relax
   \xdef\@gtempa{\write\@auxout{\string
      \newlabel{#1}{{\@currentlabel}{\thepage}}}}}\@gtempa
   \if@nobreak \ifvmode\nobreak\fi\fi\fi\@esphack}
        \gdef\@eqnlabel{#1}}
\def\@eqnlabel{}
\def\@vacuum{}
\def\draftmarginnote#1{\marginpar{\raggedright\scriptsize\tt#1}}
\def\draft{\oddsidemargin 0.0truein
        \def\@oddfoot{\sl preliminary draft \hfil
        \rm\thepage\hfil\sl\today\quad\militarytime}
        \let\@evenfoot\@oddfoot \overfullrule 3pt
        \let\label=\draftlabel
        \let\marginnote=\draftmarginnote
   \def\@eqnnum{(\theequation)\rlap{\kern\marginparsep\tt\@eqnlabel}%
\global\let\@eqnlabel\@vacuum}  }
\catcode`@=12
\begin{document}
\begin{titlepage}
\vspace*{-1cm}
\phantom{hep-ph/***}


\vskip 2.5cm
\begin{center}
{\Large\bf  SUSY Adjoint $SU(5)$ Grand Unified Model with $S_4$
Flavor Symmetry}

\end{center}
\vskip 0.2  cm
\vskip 0.5  cm
\begin{center}
{\large Gui-Jun Ding}~\footnote{e-mail address: dinggj@ustc.edu.cn}
\\
\vskip .1cm
{\it Department of Modern Physics,}
\\
{\it University of Science and Technology of China, Hefei, Anhui 230026,
China}
\\
\vskip .1cm

\end{center}
\vskip 0.7cm
\begin{abstract}
\noindent

We construct a supersymmetric (SUSY) $SU(5)$ model with the flavor
symmetry $S_4\times Z_3\times Z_4$. Three generations of adjoint
matter fields are introduced to generate the neutrino masses via the
combined type I and type III see-saw mechanism. The first two
generations of the the $\mathbf{10}$ dimensional representation in
$SU(5)$ are assigned to be a doublet of $S_4$, the second family
$\mathbf{10}$ is chose as the first component of the doublet, and
the first family as the second component. Tri-bimaximal mixing in
the neutrino sector is predicted exactly at leading order, charged
lepton mixing leads to small deviation from the tri-bimaximal mixing
pattern. Subleading contributions introduce corrections of order
$\lambda^2_c$ to all three lepton mixing angles. The model also
reproduces a realistic pattern of quark and charged lepton masses
and quark mixings. The phenomenological implications of the model
are analyzed in datail.

\end{abstract}
\end{titlepage}
\setcounter{footnote}{0}
\vskip2truecm
%
\section{Introduction}
So far there is convincing evidence that the so-called solar and
atmospheric anomaly can be well explained through the neutrino
oscillation. The mass square differences $\Delta m^2_{sol}$, $\Delta
m^2_{atm}$ and the mixing angles have been measured with good
accuracy \cite{Strumia:2006db,Schwetz:2008er,Fogli:Indication}.
Global fit to the current neutrino oscillation data demonstrates
that the observed lepton mixing matrix is remarkably compatible with
the tri-bimaximal (TB) mixing pattern \cite{TBmix}, which suggests
the following values of the mixing angles
\begin{equation}
\label{1}\sin^2\theta_{12}=\frac{1}{3},~~~\sin^2\theta_{23}=\frac{1}{2},~~\sin\theta_{13}=0
\end{equation}
This simple structure of the mixing matrix suggests that there may
be some symmetry underlying the lepton sector. Recent study
\cite{TBModel,Altarelli:2005yp,Altarelli:2005yx,Lin:2009bw,Bazzocchi:2007na,Csaki:2008qq,King:2006np,Chen:2009um,Altarelli:2009kr,Burrows:2009pi,Feruglio:2008ht,Branco:2009by,Bertuzzo:2009im,Hagedorn:2009jy,Lin:2009sq
}(also in the context of grand unified theories
\cite{Altarelli:2008bg,Morisi:2007ft}) showed that the discrete
group $A_4$ is particularly suitable to produce the TB mixing at
leading order (LO), if it is properly managed to be broken
differently in the neutrino and the charged lepton sector. In the
$A_4$ based models, it seems very difficult and unnatural to
generate the correct mass hierarchies and mixing for quarks. An
interesting solution is to enlarge the symmetry group $A_4$, two
non-Abelian discrete group $T'$ \cite{TpModel} and $S_4$
\cite{Bazzocchi:2009pv,Ding:2009iy,Altarelli:2009gn,Grimus:2009pg,Dutta:2009bj,Hagedorn:2010th,Ishimori:2008fi,Toorop:2010yh}
have been investigated. We note that both groups have a doublet
representation, which can be utilized
to give the 2+1 representation assignments for the quarks. In the
context of $U(2)$ flavor group, this assignment has been known to
give realistic quark mixing matrix and mass hierarchy \cite{u2}.
The irreducible representations of $T'$
are those of $A_4$ plus three two dimensional representations
$\mathbf{2}$, $\mathbf{2'}$ and $\mathbf{2''}$ with the
multiplication rules
$\mathbf{2}\otimes\mathbf{2}=\mathbf{2'}\otimes\mathbf{2''}=\mathbf{3}\oplus\mathbf{1}$,
$\mathbf{2}\otimes\mathbf{2'}=\mathbf{2''}\otimes\mathbf{2''}=\mathbf{3}\oplus\mathbf{1'}$
and
$\mathbf{2}\otimes\mathbf{2''}=\mathbf{2'}\otimes\mathbf{2'}=\mathbf{3}\oplus\mathbf{1''}$,
these ingredients allow us to reproduce the successful U(2)
predictions in the quark sector. By working only with the triplet
and singlet representations, $T'$ is indistinguishable from $A_4$,
thus we can replicate with $T'$ the successful construction realized
within $A_4$ in the lepton sector. $S_4$ is claimed to be the
minimal group which can predict the TB mixing in a natural way,
namely without ad hoc assumptions, from the group theory point of
view \cite{Lam:2008rs}. Actually the exact TB mixing can be realized
in the $S_4$ flavor model \cite{Bazzocchi:2009pv,Ding:2009iy}.
Moreover, the group $S_4$ as a flavor symmetry, as is shown for
example in Ref.
\cite{Dutta:2009bj,Hagedorn:2010th,Ishimori:2008fi,Toorop:2010yh},
can also give a successful description of the quark and lepton
masses and mixing angles within the framework of $SU(5)$ or $SO(10)$
grand unified theory (GUT). For a review of discrete flavor symmetry models, please see the
Refs.\cite{Altarelli:2010gt,Ishimori:2010au}.

The $SU(5)$ GUT is the simplest grand unified theory
\cite{Georgi:1974sy}, in this case each generation of the standard
model fields resides in $\overline{\mathbf 5}$ and $\mathbf{10}$
dimensional representations. To be specific, one family of
right-handed down quarks and left-handed leptons are unified in a
$\overline{\mathbf 5}$ and the rest fields of the family are in a
$\mathbf{10}$. It is well-known that neutrino masses are zero at
renormalizable level in the minimal $SU(5)$. In GUT neutrino masses
come naturally through the see-saw mechanism, where integrating out
large masses leads to the appearance of small masses. However, this
requires some extra matter fields or Higgs to be added below the GUT
scale. A popular choice is to add at least two right-handed
neutrinos which are $SU(5)$ singlets, neutrino masses are generated
through type I see-saw mechanism. The second choice is to introduce
a $\mathbf{15}$-plet of Higgs, this is the $SU(5)$ implementation of
the so-called type II see-saw mechanism. The third choice is to
generate neutrino masses through the combination of type I see-saw
with type III see-saw mechanism by adding at least one matter
multiplet in the adjoint $\mathbf{24}$ representation
\cite{Barr:2005je,Perez:2007iw,Bajc:2006ia}.

In this work, we shall construct a supersymmetric $SU(5)$ model, and
the flavor symmetry group is $S_4\times Z_3\times Z_4$, where the
auxiliary symmetry $Z_3\times Z_4$ plays an important role in
eliminating unwanted couplings, ensuring the needed vacuum alignment
and reproducing the observed charged fermion mass hierarchies. We
will introduce three generations of adjoint matter fields to
generate the neutrino masses. We remark that some variants of
$SU(5)\times S_4$ flavor models have been proposed in Refs.
\cite{Hagedorn:2010th,Ishimori:2008fi}, where three right handed
neutrinos are introduced and the neutrino masses are generated via
the type I see-saw mechanism.

The paper is organized as follows. In section 2 we discuss the
structure of the model, and the leading order (LO) results for
fermion masses and mixings are presented. In section 3 we justify
the choice of the vacuum configuration assumed in the previous
section, by minimizing the scalar potential of the theory in the
supersymmetric limit. In section 4 the subleading corrections to the
vacuum alignment and the LO results of fermion masses and flavor
mixings are discussed. In section 5 we study the phenomenological
predictions of the model in detail. Finally, section 6 is devoted to
our conclusion.

\section{The structure of the model}
In the following we present our $SU(5)$ model in the framework of
supersymmetry, which simplifies the minimization of the scalar
potential greatly. The $S_4$ group acts as a flavor symmetry of our
model, the group $S_4$ has already been studied in literature
\cite{Pakvasa:1978tx,Hagedorn:2006ug}, but with different aims and
different results. $S_4$ is the permutation group of four objects,
it has five irreducible representations $\mathbf{1_1}$,
$\mathbf{1_2}$, $\mathbf{2}$, $\mathbf{3_1}$ and $\mathbf{3_2}$, the
group theory of $S_4$ is presented in Appendix A. In addition to
$\overline{\mathbf 5}$ matter fields denoted by $F$ and the tenplet
$\mathbf{10}$ dimensional matter fields denoted by $T_{1,2,3}$, we
introduce the chiral superfields $A$ in the adjoint $\mathbf{24}$
representation \cite{Perez:2007iw}. In the Higgs sector we introduce
$H_{24}$, $H_5$ and $H_{\overline{5}}$ in order to break the gauge
symmetry $SU(5)$ into the standard model symmetry and subsequently
into the residual $SU(3)_c\times U(1)_{em}$. Moreover, $H_{45}$ and
$H_{\overline{45}}$ are introduced to avoid the wrong predictions
$M^T_d=M_{\ell}$, where $M_d$ and $M_{\ell}$ represent the mass
matrix of down type quark and charged lepton respectively. As usual,
the flavon fields are introduced to spontaneously break the $S_4$
flavor symmetry properly. The transformation rules of the matter
fields, Higgs fields and the flavon fields under $SU(5)$, $S_4$,
$Z_3$ and $Z_4$ are summarized in Table \ref{tab:trans}. The first
and the second generation of the $\mathbf{10}$ dimensional
representations are assigned to be a doublet of $S_4$, and the third
generation of $\mathbf{10}$ to $1_1$ of $S_4$. This assignment is
indicated by the heaviness of the top quark. There is the freedom of
choosing the first family or the second family as the first
component of the $S_4$ doublet. In this work, the second family
$\mathbf{10}$ is taken to be the first component of the doublet, and
the first family as the second component. If we assign the first
family $\mathbf{10}$ as the first component of the doublet and the
second family as the second component as usual, the down quark and
strange quark masses would be of the
same order\footnote{If we take $T_1$
and $T_2$ as the first and the second components of the $S_4$
doublet, i.e., $(T_1,T_2)^{T}\sim\mathbf{2}$, then
$(TF)_{\mathbf{3_1}}\sim(T_1F_2+T_2F_3,T_1F_3+T_2F_1,T_1F_1+T_2F_2)^{T}$
and
$(TF)_{\mathbf{3_2}}\sim(T_1F_2-T_2F_3,T_1F_3-T_2F_1,T_1F_1-T_2F_2)^{T}$.
We see that $T_1F_1$ and $T_2F_2$, which are related to the down and
strange quark masses respectively, appear simultaneously as the
third component of both the combinations $(TF)_{\mathbf{3_1}}$ and
$(TF)_{\mathbf{3_2}}$. The operators $TFH_{\overline{\mathbf{5}}}$
and $TFH_{\overline{\mathbf{45}}}$ combining with the flavon fields
or the composition of flavons, which transform as $\mathbf{3_1}$ or
$\mathbf{3_2}$, contribute to the first two families down quark and
charged lepton masses after the $S_4$ and GUT symmetry breaking. As
a result, the down and strange quark messes would be of the same order
except for the case that the $(TF)_{\mathbf{3_1}}$ and
$(TF)_{\mathbf{3_2}}$ relevant contributions to down quark or
strange quark cancel with each other. We notice that the same view
has been put forward by Altarelli et al. \cite{Altarelli:2010gt}. Whereas if we choose $(T_2,T_1)^{T}\sim\mathbf{2}$ as we proposed,  then
$(TF)_{\mathbf{3_1}}\sim(T_2F_2+T_1F_3,T_2F_3+T_1F_1,T_2F_1+T_1F_2)^{T}$
and
$(TF)_{\mathbf{3_2}}\sim(T_2F_2-T_1F_3,T_2F_3-T_1F_1,T_2F_1-T_1F_2)^{T}$, the degeneracy between first two families down quark (charged lepton) masses is dissolved.}
without fine tuning unless some special mechanisms are introduced
such as Ref. \cite{Hagedorn:2010th}. Three generation of
$\overline{\mathbf{5}}$ fields $F$ are assigned to $\mathbf{3_1}$ of
$S_4$, and three generations of adjoint matter fields $A$ are also
assigned to be $\mathbf{3_1}$ of $S_4$. Fermion masses and mixings arise from the spontaneous breaking of the
flavor symmetry by means of the flavon fields. In the following, we shall
discuss the LO predictions for fermion masses and flavor mixings.
For the time being we assume that the scalar components of the
flavon fields acquire vacuum expectation values (VEV) according to
the following scheme
\begin{eqnarray}
\nonumber&&\langle\chi\rangle=\left(\begin{array}{c}v_{\chi}\\v_{\chi}\\v_{\chi}\end{array}\right),~~~~~~~\langle\varphi\rangle=\left(\begin{array}{c}v_{\varphi}\\
v_{\varphi}\end{array}\right),~~~~~~~\langle\zeta\rangle=0\\
\nonumber&&\langle\phi\rangle=\left(\begin{array}{c}0\\
v_{\phi}\\0\end{array}\right),~~~~~~~\langle\eta\rangle=\left(\begin{array}{c}0\\v_{\eta}\end{array}\right)\\
\label{2}&&\langle\Delta\rangle=\left(\begin{array}{c}v_{\Delta}\\0\\0\end{array}\right),~~~~~~~\langle\xi\rangle=v_{\xi}
\end{eqnarray}
We will prove it to be a natural solution of the minimization of the scalar potential in section 3. Furthermore
we take the VEVs (scaled by the cutoff $\Lambda$)
$v_{\chi}/\Lambda$, $v_{\varphi}/\Lambda$, $v_{\phi}/\Lambda$,
$v_{\eta}/\Lambda$, $v_{\Delta}/\Lambda$ and $v_{\xi}/\Lambda$ to be
of the same order of magnitude about ${\cal O}(\lambda^2_c)$ with
$\lambda_c\simeq0.22$ being the Cabibbo angle, and we will
parameterize the ratio ${\rm VEV}/\Lambda$ by the parameter
$\varepsilon$. This order of magnitude is indicated by the observed
ratios of up quarks and down quarks and charged lepton masses, by
the scale of the light neutrino masses and is also compatible with
the current bounds on the deviations from TB mixing for the leptons. We note that the assumed size of the VEVs can be partially explained by the minimization of the scalar potential, as it will be clearer in the following.
\begin{table}[hptb]
\begin{center}
\begin{tabular}{|c|c|c|c|c||c|c|c|c|c||c|c|c|c|c|c|c|}\hline\hline
Fields& $T_3$  & $(T_2,T_1)^{T}$ & $F$ &  $A$ &$H_5$ & $H_{45}$ &
$H_{\overline{5}}$ &  $H_{\overline{45}}$&$H_{24}$&$\chi$ &
$\varphi$ & $\zeta$& $\phi$ & $\eta$ & $\Delta$& $\xi$\\ \hline

$SU(5)$&$\mathbf{10}$&$\mathbf{10}$&$\overline{\mathbf{5}}$
&$\mathbf{24}$ & $\mathbf{5}$& $\mathbf{45}$&$\overline{\mathbf{5}}$
&$\overline{\mathbf{45}}$ &$\mathbf{24}$ & $\mathbf{1}$&
$\mathbf{1}$ & $\mathbf{1}$ &$\mathbf{1}$ & $\mathbf{1}$& $\mathbf{1}$&
$\mathbf{1}$\\ \hline

$\rm{S_4}$& $\mathbf{1_1}$& $\mathbf{2}$ &
$\mathbf{3_1}$&$\mathbf{3_1}$ & $\mathbf{1_1}$ & $\mathbf{1_1}$ &
$\mathbf{1_1}$ & $\mathbf{1_1}$&$\mathbf{1_1}$ & $\mathbf{3_1}$& $\mathbf{2}$ & $\mathbf{1_2}$& $\mathbf{3_1}$ & $\mathbf{2}$ & $\mathbf{3_1}$ & $\mathbf{1_1}$ \\
\hline

$\rm{Z_{3}}$& 1 & $\omega$& 1& 1  & 1 & 1 & $\omega$ & 1& 1& 1  &1 & $\mathbf{1}$&
$\omega^2$ & $\omega^2$ & $\omega$& $\omega$\\ \hline

$\rm{Z_{4}}$&1  & $i$& $-i$ & $i$  & 1 & -1 & $1$ & -1&  1 &-1 &-1 & $-1$&
$i$ & $i$ & $i$ & $i$ \\ \hline

$U(1)_R$&1  & 1& 1 & 1  & 0 &0 & $0$ &0&  0 &0 &0 &0 &0  & 0 & 0 &0
\\\hline\hline
\end{tabular}
\end{center}
\caption{\label{tab:trans}Fields and their transformation properties
under the symmetry groups $SU(5)$, $S_4$, $Z_3$ and $Z_4$, where
$\omega=e^{i2\pi/3}=(-1+i\sqrt{3})/2$.}
\end{table}
\subsection{Neutrino sector}
\begin{figure}
\begin{center}
\begin{tabular}{cc}
\includegraphics[scale=.745]{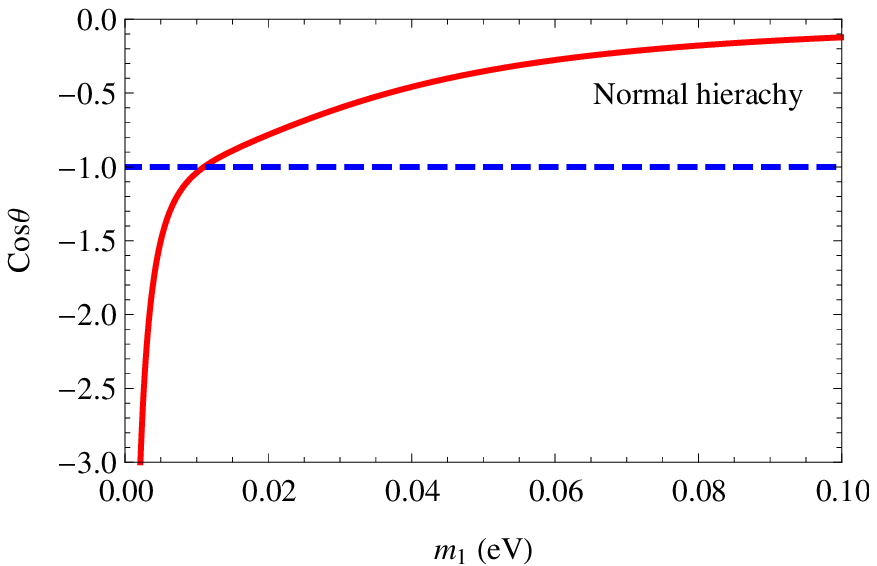}&\includegraphics[scale=.745]{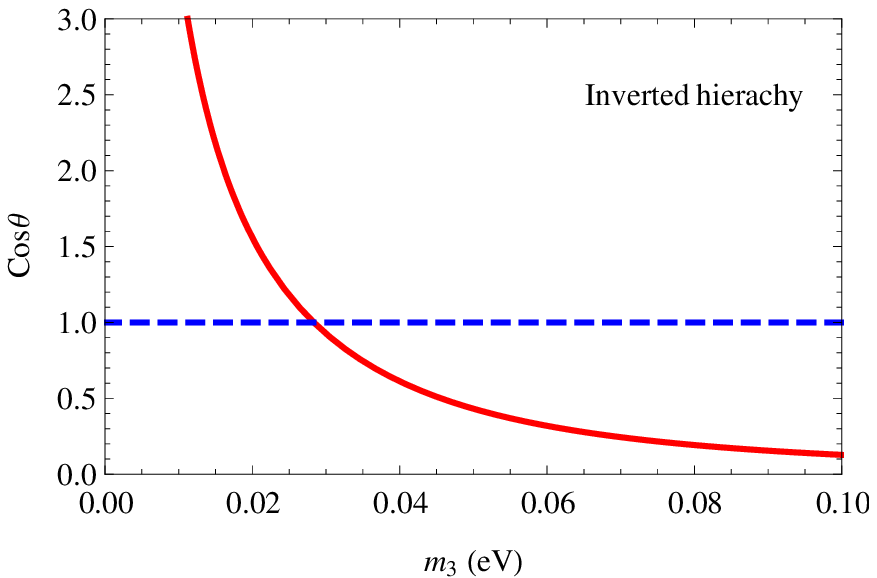}\\
(a)&(b)
\end{tabular}
\caption{\label{fig:cosphi} $\cos\theta$ as a function of the
lightest neutrino mass $m_l$ for both normal hierarchy and inverted
hierarchy spectrum, where we take $\Delta
m^2_{sol}=7.67\times10^{-5}{\rm eV^2}$ and $\Delta
m^2_{atm}=2.46(2.45)\times10^{-3}{\rm eV^2}$ for normal hierarchy
(inverted hierarchy).}
\end{center}
\end{figure}

The LO superpotential which contributes to the neutrino masses is
given by
\begin{equation}
\label{22}w_{\nu}=y_{\nu}(FA)_{1_1}H_5+\lambda_1(AA)_{3_1}\chi+\lambda_2(AA)_2\varphi
\end{equation}
The adjoint matter field $A$ decomposes under the standard model as
$A=(\rho_8,\rho_3,\rho_{(3,2)},\rho_{(\bar{3},2)},\rho_0)=(8,1,0)\oplus(1,3,0)\oplus(3,2,-5/6)\oplus(\bar{3},2,5/6)\oplus(1,1,0)$,
obviously there are both $SU(2)$ triplet $\rho_3$ and singlet
$\rho_0$ with hypercharge $Y=0$ in the model. The neutrino masses
are generated through the type I (mediated by the SU(2) singlet
$\rho_0$ of $A$) and type III (mediated by the SU(2) triplet
$\rho_3$ of $A$) see-saw mechanism. The Dirac mass matrices is
obtained from the first term in Eq.(\ref{22}),
\begin{equation}
\label{23}M^D_{\rho_3}=\frac{1}{2}y_{\nu}v_5\left(\begin{array}{ccc}1&0&0\\
0&0&1\\
0&1&0
\end{array}\right),~~~~M^{D}_{\rho_0}=\frac{\sqrt{15}}{10}y_{\nu}v_5\left(\begin{array}{ccc}
1&0&0\\
0&0&1\\
0&1&0
\end{array}\right)
\end{equation}
where $M^D_{\rho_3}$ and $M^D_{\rho_0}$ are the Dirac mass matrices
associated with $\rho_3$ and $\rho_0$ respectively. The last two
terms in Eq.(\ref{22}) lead to the Majorana mass matrices of
$\rho_3$ and $\rho_0$
\begin{eqnarray}
\nonumber&&M^{M}_{\rho_3}=\left(\begin{array}{ccc}2\lambda_1v_{\chi}&-\lambda_1v_{\chi}+\lambda_2v_{\varphi}&-\lambda_1v_{\chi}+\lambda_2v_{\varphi}\\
-\lambda_1v_{\chi}+\lambda_2v_{\varphi}&2\lambda_1v_{\chi}+\lambda_2v_{\varphi}&-\lambda_1v_{\chi}\\
-\lambda_1v_{\chi}+\lambda_2v_{\varphi}&-\lambda_1v_{\chi}&2\lambda_1v_{\chi}+\lambda_2v_{\varphi}
\end{array}\right)\\
\label{24}&&M^{M}_{\rho_0}=M^{M}_{\rho_3}
\end{eqnarray}
It is notable that the Majorana mass matrices of $\rho_3$ and
$\rho_0$ are exactly the same. As a result, the mass spectrums of
$\rho_3$ and $\rho_0$ are degenerate. This degeneracy is violated at
NLO by the Higgs $H_{24}$. We note the VEVs $\langle\chi\rangle$ and
$\langle\varphi\rangle$ are invariant under the action of $S_4$
elements $TSTS^2$, $TST$ and $S^2$, consequently the $S_4$ flavor
symmetry is broken down to the Klein four subgroup in the neutrino
sector. The light neutrino mass matrix is the sum of type I and type
III see-saw contributions
\begin{eqnarray}
\nonumber M_{\nu}&=&-(M^{D}_{\rho3})^T(M^{M}_{\rho_3})^{-1}M^D_{\rho_3}-(M^{D}_{\rho0})^T(M^{M}_{\rho_0})^{-1}M^D_{\rho_0}\\
\label{25}&=&\left(\begin{array}{ccc}\frac{-a-b}{5b(3a-b)}&\frac{-a+b}{5b(3a-b)}&\frac{-a+b}{5b(3a-b)}\\
\frac{-a+b}{5b(3a-b)}&\frac{-3a^2-4ab+b^2}{5b(9a^2-b^2)}&\frac{-3a^2+2ab-b^2}{5b(9a^2-b^2)}\\
\frac{-a+b}{5b(3a-b)}&\frac{-3a^2+2ab-b^2}{5b(9a^2-b^2)}&\frac{-3a^2-4ab+b^2}{5b(9a^2-b^2)}
\end{array}\right)y^2_{\nu}v^2_5
\end{eqnarray}
where
\begin{equation}
\label{26}a\equiv\lambda_1v_{\chi},~~~~b\equiv\lambda_2v_{\varphi}
\end{equation}
The above light neutrino mass matrix $M_{\nu}$ is
$2\leftrightarrow3$ invariant and it satisfies the magic symmetry
$(M_{\nu})_{11}+(M_{\nu})_{13}=(M_{\nu})_{22}+(M_{\nu})_{32}$.
Therefore it is exactly diagonalized by the TB mixing matrix
\begin{equation}
\label{27}U^{T}_{\nu}M_{\nu}U_{\nu}={\rm diag}(m_1,m_2,m_3)
\end{equation}
where $m_{1,2,3}$ are the light neutrino masses, in unit of
$\frac{2}{5}y^2_{\nu}v^2_5$ they are
\begin{eqnarray}
\nonumber&&m_1=\frac{1}{|3a-b|}\\
\nonumber&&m_2=\frac{1}{2|b|}\\
\label{28}&&m_3=\frac{1}{|3a+b|}
\end{eqnarray}
The neutrino mass spectrum can be normal hierarchy (NH) or inverted
hierarchy (IH). The unitary matrix $U_{\nu}$ is given by
\begin{equation}
\label{29}U_{\nu}=U_{TB}\,{\rm
diag}(e^{-i\alpha_1/2},e^{-i\alpha_2/2},e^{-i\alpha_3/2})
\end{equation}
$U_{TB}$ is the well-known TB mixing matrix
\begin{equation}
\label{30}U_{TB}=\left(\begin{array}{ccc}\sqrt{\frac{2}{3}}&\frac{1}{\sqrt{3}}&0\\
-\frac{1}{\sqrt{6}}&\frac{1}{\sqrt{3}}&\frac{1}{\sqrt{2}}\\
-\frac{1}{\sqrt{6}}&\frac{1}{\sqrt{3}}&-\frac{1}{\sqrt{2}}
\end{array}\right)
\end{equation}
The phases $\alpha_1$, $\alpha_2$ and $\alpha_3$ are
\begin{eqnarray}
\nonumber&&\alpha_1={\rm
arg}(-{y^2_{\nu}v^2_5}/{(3a-b)})\\
\nonumber&&\alpha_2={\rm arg}(-{y^2_{\nu}v^2_5}/{b})\\
\label{31}&&\alpha_3={\rm arg}(-{y^2_{\nu}v^2_5}/{(3a+b)})
\end{eqnarray}
According to Eq.(\ref{28}), the light neutrino mass spectrum is
directly related to the heavy neutrino ($\rho_3$ or $\rho_0$)
masses, it is determined by only two complex parameters $a$ and $b$. In the following, we shall follow the method of Ref. \cite{Bertuzzo:2009im} to analyze the light neutrino mass spectrum in detail.
For the sake of convenience, we define
\begin{equation}
\label{32}\frac{a}{b}=re^{i\theta}
\end{equation}
Note that the parameter $r$ is real and positive and the phase
$\theta$ is between 0 and $2\pi$. We can express $r$ and the phase
$\theta$ in term of light neutrino masses as follows
\begin{eqnarray}
\label{33}&&r=\frac{1}{3}\sqrt{\frac{2m^2_2}{m^2_1}+\frac{2m^2_2}{m^2_3}-1}\\
\label{34}&&\cos\theta=\frac{\frac{m^2_2}{m^2_3}-\frac{m^2_2}{m^2_1}}{\sqrt{\frac{2m^2_2}{m^2_1}+\frac{2m^2_2}{m^2_3}-1}}
\end{eqnarray}
Experimentally, only the mass square differences have been measured.
For normal (inverted) hierarchy they are
\begin{eqnarray}\
\nonumber&&\Delta
m^2_{sol}=m^2_2-m^2_1=(7.67^{+0.22}_{-0.21})\times10^{-5}{\rm eV}^2\\
\label{35}&&\Delta
m^2_{atm}=|m^2_3-m^2_1(m^2_2)|=(2.46(2.45)\pm0.15)\times10^{-3}{\rm
eV}^2
\end{eqnarray}
By expressing the neutrino masses in terms of the lightest neutrino
mass $m_l$ ($m_l=m_1$ for NH and $m_3$ for IH), $\Delta m^2_{sol}$
and $\Delta m^2_{atm}$, $\cos\theta$ becomes a function of the
lightest neutrino mass $m_l$. We display $\cos\theta$ versus $m_l$
in Fig.\ref{fig:cosphi} for both normal hierarchy and inverted
hierarchy spectrum. Imposing the condition $|\cos\theta|\leq1$, we
obtain the following constraints on the lightest neutrino mass
\begin{eqnarray}
\nonumber&& m_1\geq0.011{\rm eV},~~~ {\rm for~ normal~ hierarchy}\\
\label{36}&& m_3\geq0.028{\rm eV},~~~ {\rm for ~inverted~ hierarchy}
\end{eqnarray}
The results presented so far are of course approximate since the
model gets corrections when higher dimensional operators are
included in the Lagrangian. If the mixing of the left handed charged
lepton is neglected, we can straightforwardly derive the
neutrinoless double decay parameters $|m_{ee}|$
\begin{equation}
\label{37}|m_{ee}|=\frac{m_1}{3}\sqrt{2-\frac{m^2_2}{m^2_1}+2\frac{m^2_2}{m^2_3}}
\end{equation}
\subsection{Up quarks sector}
The LO superpotential invariant under symmetry group $SU(5)\times
S_4\times Z_3\times Z_4$, which gives rise to the masses of the up
type quarks after $S_4$ and $SU(5)$ symmetry breaking, is given by
\begin{eqnarray}
\nonumber
w_u&=&y_tT_3T_3H_5+\sum^4_{i=1}\frac{y_{ci}}{\Lambda^2}TT{\cal
O}^{(1)}_iH_5+\frac{y_{ut1}}{\Lambda^2}TT_3(\phi\chi)_2H_5+
\frac{y_{ut2}}{\Lambda^2}TT_3(\eta\varphi)_2H_5\\
\label{3}&&+\frac{y_{ut3}}{\Lambda^2}TT_3\eta\zeta H_5+\frac{y_{ct}}{\Lambda}TT_3\eta
H_{45}
\end{eqnarray}
where
\begin{eqnarray}
\nonumber {\cal
O}^{(1)}&=&\{(\phi\phi)_{1_1},(\phi\phi)_2,(\eta\eta)_{1_1},(\eta\eta)_2\}
\end{eqnarray}
With the vacuum alignment in Eq.(\ref{2}), it is immediate
to derive the mass matrix as follows
\begin{equation}
\label{4}M_u=\left(\begin{array}{ccc}
0&0&4(y_{ut1}\frac{v_{\phi}v_{\chi}}{\Lambda^2}+y_{ut2}\frac{v_{\eta}v_{\varphi}}{\Lambda^2})v_5\\
0&8(y_{c2}\frac{v^2_{\phi}}{\Lambda^2}+y_{c4}\frac{v^2_{\eta}}{\Lambda^2})v_5&8y_{ct}\frac{v_{\eta}}{\Lambda}v_{45}+4y_{ut1}\frac{v_{\phi}v_{\chi}}{\Lambda^2}v_5\\
4(y_{ut1}\frac{v_{\phi}v_{\chi}}{\Lambda^2}+y_{ut2}\frac{v_{\eta}v_{\varphi}}{\Lambda^2})v_5&-8y_{ct}\frac{v_{\eta}}{\Lambda}v_{45}+4y_{ut1}\frac{v_{\phi}v_{\chi}}{\Lambda^2}v_5&8y_tv_5
\end{array}\right)
\end{equation}
We note that the Higgs field $H_5$ ($H_{45}$) induces a symmetric
(antisymmetric) contribution to $M_u$. Since the product
decompositions $(TT)_{1_1}\sim T_1T_2+T_2T_1$ and
$(TT)_2\sim\left(\begin{array}{c}T^2_1\\T^2_2\end{array}\right)$ are
both symmetric under the exchange of the two tenplets, the combinations of $T_3T_3H_{45}$ and $TTH_{45}$ with arbitrary number of flavon fields
don't contribute to up quarks masses. Therefore the corresponding terms are omitted from the beginning. The mass matrix $M_u$
is diagonalized by bi-unitary transformation
\begin{equation}
\label{5}V^{u\dagger}_{R}M_uV^{u}_{L}={\rm diag}(m_u,m_c,m_t)
\end{equation}
The up type quark masses are given by
\begin{eqnarray}
\nonumber&&m_u\simeq\Big|\frac{2(y_{c2}v^2_{\phi}+y_{c4}v^2_{\eta})(y_{ut1}v_{\phi}v_{\chi}/\Lambda^2+y_{ut2}v_{\eta}v_{\varphi}/\Lambda^2)^2}{y_t(y_{c2}v^2_{\phi}+y_{c4}v^2_{\eta})v^2_5+y^2_{ct}v^2_{\eta}v^2_{45}}v^3_5\Big|\\
\nonumber&&m_c\simeq\Big|8(y_{c2}\frac{v^2_{\phi}}{\Lambda^2}+y_{c4}\frac{v^2_{\eta}}{\Lambda^2})v_5+8\frac{y^2_{ct}}{y_t}\frac{v^2_{\eta}}{\Lambda^2}\frac{v^2_{45}}{v_5}\Big|\\
\label{6}&&m_t\simeq\Big|8y_tv_5\Big|
\end{eqnarray}
The mixing matrix $V^{u}_{L}$ is
\begin{equation}
\label{7}V^{u}_{L}\simeq\left(\begin{array}{ccc}
1&s^{u}_{12}&s^{u}_{13}\\
-s^{u*}_{12}&1&s^{u}_{23}\\
-s^{u*}_{13}+s^{u*}_{12}s^{u*}_{23}&-s^{u*}_{23}&1
\end{array}\right)
\end{equation}
where
\begin{eqnarray}
\nonumber&&s^{u}_{12}=-\frac{1}{2}\Big[\frac{y_{ct}(y_{ut1}v_{\phi}v_{\chi}+y_{ut2}v_{\eta}v_{\varphi})v_5v_{45}}{y_t(y_{c2}v^2_{\phi}+y_{c4}v^2_{\eta})v^2_5+y^2_{ct}v^2_{\eta}v^2_{45}}\frac{v_{\eta}}{\Lambda}\Big]^{*}\\
\nonumber&&s^{u}_{23}=\Big(-\frac{y_{ct}}{y_t}\frac{v_{45}}{v_5}\frac{v_{\eta}}{\Lambda}+\frac{y_{ut1}}{2y_{t}}\frac{v_{\phi}v_{\chi}}{\Lambda^2}\Big)^{*}\\
\label{8}&&s^{u}_{13}=\frac{1}{2}\Big(\frac{y_{ut1}}{y_t}\frac{v_{\phi}v_{\chi}}{\Lambda^2}+\frac{y_{ut2}}{y_t}\frac{v_{\eta}v_{\varphi}}{\Lambda^2}\Big)^{*}
\end{eqnarray}
We note that there is a mixing of order $\lambda^2_c$ between the
first and the second family, although the (12) and $(21)$ elements
of $M_u$ vanish at LO. The top quark mass is generated at tree
level, and the mass hierarchies among the up quarks are reproduced
naturally given the VEVs $v_{\chi}$, $v_{\varphi}$, $v_{\phi}$ and
$v_{\eta}$ of order $\lambda^2_c\Lambda$.
\subsection{Down type quarks and charged leptons sector}
The superpotential generating the masses of down quarks and charged
lepton is
\begin{eqnarray}
\nonumber w_d&=&\frac{y_b}{\Lambda}T_3F\phi
H_{\overline{5}}+\frac{y_{s1}}{\Lambda^2}(TF)_{3_1}(\Delta\Delta)_{3_1}H_{\overline{45}}+\frac{y_{s2}}{\Lambda^2}(TF)_{3_1}\Delta\xi
H_{\overline{45}}+\sum^9_{i=1}\frac{y_{di}}{\Lambda^3}T_3F{\cal
O}^{(2)}_iH_{\overline{5}}\\
\label{9}&&+\sum^{6}_{i=1}\frac{x_{di}}{\Lambda^3}T_3F{\cal
O}^{(3)}_iH_{\overline{45}}+\sum^{7}_{i=1}\frac{z_{di}}{\Lambda^3}TF{\cal
O}^{(4)}_iH_{\overline{5}}+...
\end{eqnarray}
where dots stand for higher dimensional operators.
\begin{eqnarray}
\nonumber &&{\cal
O}^{(2)}=\{\chi^2\phi,\chi^2\eta,\varphi\chi\phi,\varphi\chi\eta,\varphi^2\phi,\chi\phi\zeta,\chi\eta\zeta,\varphi\phi\zeta,\phi\zeta^2\}\\
\nonumber&&{\cal
O}^{(3)}=\{\phi^3,\phi^2\eta,\phi\eta^2,\Delta^3,\Delta^2\xi,\Delta\xi^2\}\\
\label{10}&&{\cal
O}^{(4)}=\{\phi^2\chi,\phi^2\varphi,\phi^2\zeta,\eta\phi\chi,\eta\phi\varphi,\eta\phi\zeta,\eta^2\chi\}
\end{eqnarray}
For the last three terms in Eq.(\ref{9}), one operator frequently
induces several different contractions, we should take into account
all possible independent contractions for each operator. Note that
the auxiliary symmetry $Z_3\times Z_4$ imposes different powers of
the flavon fields for the bottom (tau), strange (muon) and down
quark (electron) mass relevant terms. In the expansion in powers of
$1/\Lambda$, the bottom (tau) mass is generated at order
$1/\Lambda$, the strange (muon) and down quark (electron) masses are
generated at order $1/\Lambda^2$ and $1/\Lambda^3$ respectively. As
a result, if we only consider the LO operators suppressed by
$1/\Lambda$ and $1/\Lambda^2$, the down quark and electron would be
massless.

Recalling the vacuum configuration in Eq.(\ref{2}), we can write
down the mass matrix for down quarks and charged leptons as follows
\begin{eqnarray}
\label{11}&&M_d=\left(\begin{array}{ccc}y^{d}_{11}\varepsilon^3v_{\overline{5}}&y^d_{12}\varepsilon^3v_{\overline{5}}&y^d_{13}\varepsilon^3v_{\overline{5}}+2y^{d'}_{13}\varepsilon^3v_{\overline{45}}\\
y^d_{21}\varepsilon^3v_{\overline{5}}&2y^d_{22}\varepsilon^2v_{\overline{45}}+y^{d'}_{22}\varepsilon^3v_{\overline{5}}&y^d_{23}\varepsilon^3v_{\overline{5}}\\
2y^d_{22}\varepsilon^2v_{\overline{45}}+y^{d'}_{31}\varepsilon^3v_{\overline{5}}&y^d_{32}\varepsilon^3v_{\overline{5}}&y^d_{33}\varepsilon
v_{\overline{5}}
\end{array}\right)\\
\label{12}&&M_{\ell}=\left(\begin{array}{ccc}y^d_{11}\varepsilon^3v_{\overline{5}}&y^{d}_{21}\varepsilon^3v_{\overline{5}}&-6y^d_{22}\varepsilon^2v_{\overline{45}}+y^{d'}_{31}\varepsilon^3v_{\overline{5}}\\
y^d_{12}\varepsilon^3v_{\overline{5}}&-6y^d_{22}\varepsilon^2v_{\overline{45}}+y^{d'}_{22}\varepsilon^3v_{\overline{5}}&y^d_{32}\varepsilon^3v_{\overline{5}}\\
y^d_{13}\varepsilon^3v_{\overline{5}}-6y^{d'}_{13}\varepsilon^3v_{\overline{45}}&y^d_{23}\varepsilon^3v_{\overline{5}}&y^d_{33}\varepsilon
v_{\overline{5}}
\end{array}\right)
\end{eqnarray}
where the coefficients $y^d_{ij}$ ($i,j=1,2,3$), $y^{d'}_{13}$,
$y^{d'}_{22}$ and $y^{d'}_{31}$ are linear combinations of the
leading order coefficients, $y^d_{33}$ coincides with the LO
parameter $y_b$ up to corrections of order $\varepsilon^2$ which
origins from the operators $T_3F{\cal O}^{(3)}H_{\overline{5}}$. The
factor of 3 difference in the $(13)$, $(22)$ and $(31)$ elements
between $M_d$ and $M_{\ell}$ is the so-called Georgi-Jarlskog factor
\cite{Georgi:1979df}, which is induced by the Higgs
$H_{\overline{45}}$. Similar to the up type quarks, the mass
matrices $M_d$ and $M_{\ell}$ can be diagonalized by the following
transformations
\begin{equation}
\label{13}V^{d\dagger}_RM_dV^d_L={\rm
diag}(m_d,m_s,m_b),~~~~V^{\ell\dagger}_RM_{\ell}V^{\ell}_L={\rm
diag}(m_e,m_{\mu},m_{\tau})
\end{equation}
The mass eigenvalues are given by
\begin{eqnarray}
\nonumber&&m_d\simeq|y^d_{11}\varepsilon^3v_{\overline{5}}-y^d_{12}y^d_{21}\varepsilon^4v^2_{\overline{5}}/(2y^d_{22}v_{\overline{45}})-2y^d_{13}y^d_{22}\varepsilon^4v_{\overline{45}}/y^d_{33}-4y^{d'}_{13}y^d_{22}\varepsilon^4v^2_{\overline{45}}/(y^d_{33}v_{\overline{5}})|\\
\nonumber&&m_s\simeq|2y^d_{22}\varepsilon^2v_{\overline{45}}+y^{d'}_{22}\varepsilon^3v_{\overline{5}}|\\
\label{14}&&m_b\simeq|y^d_{33}\varepsilon v_{\overline{5}}|
\end{eqnarray}
and
\begin{eqnarray}
\nonumber&&m_e\simeq|y^d_{11}\varepsilon^3v_{\overline{5}}+
{y^d_{12}y^d_{21}}\varepsilon^4v^2_{\overline{5}}/({6y^d_{22}}v_{\overline{45}})
+6y^d_{13}{y^d_{22}\varepsilon^4v_{\overline{45}}}/{y^d_{33}}
-36{y^{d'}_{13}y^d_{22}\varepsilon^4v^2_{\overline{45}}}/({y^d_{33}}v_{\overline{5}})|\\
\nonumber&&m_{\mu}\simeq|-6y^d_{22}\varepsilon^2v_{\overline{45}}+y^{d'}_{22}\varepsilon^3v_{\overline{5}}|\\
\label{15}&&m_{\tau}\simeq|y^d_{33}\varepsilon v_{\overline{5}}|
\end{eqnarray}
The diagonalization matrices $V^{d}_{L}$ and $V^{\ell}_L$ are
\begin{equation}
\label{16}V^{d}_L\simeq\left(\begin{array}{ccc}
1&(\frac{y^d_{21}}{2y^d_{22}}\frac{v_{\overline{5}}}{v_{\overline{45}}}\varepsilon)^{*}&(2\frac{y^d_{22}}{y^d_{33}}\frac{v_{\overline{45}}}{v_{\overline{5}}}\varepsilon+\frac{y^{d'}_{31}}{y^d_{33}}\varepsilon^2)^{*}\\
-\frac{y^d_{21}}{2y^d_{22}}\frac{v_{\overline{5}}}{v_{\overline{45}}}\varepsilon&1&(\frac{y^d_{32}}{y^d_{33}}\varepsilon^2)^{*}\\
-2\frac{y^d_{22}}{y^d_{33}}\frac{v_{\overline{45}}}{v_{\overline{5}}}\varepsilon-\frac{y^{d'}_{31}}{y^d_{33}}\varepsilon^2&-\frac{y^d_{32}}{y^d_{33}}\varepsilon^2-\frac{y^{d*}_{21}y^d_{22}}{y^{d*}_{22}y^d_{33}}|\varepsilon|^2&1
\end{array}\right)
\end{equation}
\begin{equation}
\label{17}V^{\ell}_L\simeq\left(\begin{array}{ccc}
1&-(\frac{y^d_{12}}{6y^d_{22}}\frac{v_{\overline{5}}}{v_{\overline{45}}}\varepsilon)^{*}&(\frac{y^{d}_{13}}{y^d_{33}}\varepsilon^2-6\frac{y^{d'}_{13}}{y^d_{33}}\frac{v_{\overline{45}}}{v_{\overline{5}}}\varepsilon^2)^{*}\\
\frac{y^d_{12}}{6y^d_{22}}\frac{v_{\overline{5}}}{v_{\overline{45}}}\varepsilon&1&(\frac{y^d_{23}}{y^d_{33}}\varepsilon^2)^*\\
-\frac{y^{d}_{13}}{y^d_{33}}\varepsilon^2+6\frac{y^{d'}_{13}}{y^d_{33}}\frac{v_{\overline{45}}}{v_{\overline{5}}}\varepsilon^2&-\frac{y^d_{23}}{y^d_{33}}\varepsilon^2&1
\end{array}\right)
\end{equation}
As is shown in Eq.(\ref{14}) and Eq.(\ref{15}), obviously we have
\begin{equation}
\label{18}m_{\tau}\simeq m_{b},~~~m_{\mu}\simeq3m_{s}
\end{equation}
The well-known bottom-tau unification and the Georgi-Jarlskog relation \cite{Georgi:1979df} between the down type quark and the charged lepton masses for the second generation are produced in the present model. We note that generally the vanishing of the (11) elements of both the down quark and charged lepton mass matrices is required, to obtain the Georgi-Jarlskog relation for both the first and second generations simultaneously.
The mass difference of electron and down quark is induced by the Higgs field
$H_{\overline{45}}$, acceptable values of the masses for electron
and down quark can be accomodated due to the Georgi-Jarlskog factor.
It is well-known that the CKM mixing between the first and the
second family is exactly described by the Cabibbo angle. In order to
satisfy this phenomenological constraint, for the parameters
$y^d_{21}$ and $y^d_{22}$ of order ${\cal O}(1)$ we could choose
\begin{equation}
\label{19}v_{\overline{45}}\sim\lambda_cv_{\overline{5}}
\end{equation}
Furthermore we can see from Eq.(\ref{16}) and Eq.(\ref{17}) that the
mixing angle between the first and the second family charged leptons
approximately is $\lambda_c/3$.
The resulting quark mixing matrix is given by
\begin{equation}
\label{20}V_{CKM}=V^{u\dagger}_LV^d_L
\end{equation}
We can straightforwardly read the CKM matrix elements as follows
\begin{eqnarray}
\nonumber&&V_{ud}\simeq V_{cs}\simeq V_{tb}\simeq1\\
\nonumber&&V^{*}_{us}\simeq
-V_{cd}\simeq\frac{y^d_{21}}{2y^d_{22}}\frac{v_{\overline{5}}}{v_{\overline{45}}}\varepsilon+\frac{1}{2}\frac{y_{ct}(y_{ut1}v_{\phi}v_{\chi}+y_{ut2}v_{\eta}v_{\varphi})v_5v_{45}}{y_t(y_{c2}v^2_{\phi}+y_{c4}v^2_{\eta})v^2_5+y^2_{ct}v^2_{\eta}v^2_{45}}\frac{v_{\eta}}{\Lambda}\\
\nonumber&&V^*_{ub}=2\frac{y^d_{22}}{y^d_{33}}\frac{v_{\overline{45}}}{v_{\overline{5}}}\varepsilon+\frac{y^{d'}_{31}}{y^d_{33}}\varepsilon^2-\frac{y_{ut1}}{2y_t}\frac{v_{\phi}v_{\chi}}{\Lambda^2}-\frac{y_{ut2}}{2y_t}\frac{v_{\eta}v_{\varphi}}{\Lambda^2}+\frac{1}{2}\frac{y^2_{ct}(y_{ut1}v_{\phi}v_{\chi}+y_{ut2}v_{\eta}v_{\varphi})v^2_{45}}{y^2_t(y_{c2}v^2_{\phi}+y_{c4}v^2_{\eta})v^2_5+y_ty^2_{ct}v^2_{\eta}v^2_{45}}\frac{v^2_{\eta}}{\Lambda^2}\\
\nonumber&&V^{*}_{cb}\simeq-V_{ts}\simeq\frac{y_{ct}v_{45}}{y_tv_5}\frac{v_{\eta}}{\Lambda}\\
\label{21}&&V_{td}=-2\frac{y^d_{22}}{y^d_{33}}\frac{v_{\overline{45}}}{v_{\overline{5}}}\varepsilon-\frac{y^{d'}_{31}}{y^d_{33}}\varepsilon^2+\frac{y_{ut1}}{2y_t}\frac{v_{\phi}v_{\chi}}{\Lambda^2}+\frac{y_{ut2}}{2y_t}\frac{v_{\eta}v_{\varphi}}{\Lambda^2}+\frac{y_{ct}y^d_{21}}{2y_ty^d_{22}}\frac{v_{45}}{v_5}\frac{v_{\overline{5}}}{v_{\overline{45}}}\frac{v_{\eta}}{\Lambda}\varepsilon
\end{eqnarray}
We note that the CKM elements $V_{us}$, $V_{cd}$, $V_{ub}$ and
$V_{td}$ are dominantly determined by the mixing in the down type
quark sector, $V_{cb}$ and $V_{ts}$ origin from the left handed up
quarks mixing. Considering
$v_{\overline{45}}/v_{\overline{5}}\sim\lambda_c$, we find that
$V_{us}$ and $V_{cd}$ are of order $\lambda_c$, $V_{cb}$ and
$V_{ts}$ of order $\lambda^2_c$, $V_{ub}$ and $V_{td}$ are of order
$\lambda^3_c$. The correct pattern of CKM mixing matrix is
reproduced.

Due to the non-trivial mixing $V^{\ell}_L$ present in the charged
lepton sector, we note that the lepton mixing is not the TB mixing,
although the light neutrino mass matrix is exactly diagonalized by
the TB mixing matrix. The lepton mixing matrix (PMNS matrix) is
given by
\begin{equation}
\label{38}U_{PMNS}=V^{\ell\dagger}_LU_{\nu}
\end{equation}
Consequently the lepton mixing angles are
\begin{eqnarray}
\nonumber&&\sin\theta_{13}=|(U_{PMNS})_{e3}|\simeq\Big|\frac{y^d_{12}}{6\sqrt{2}y^d_{22}}\frac{v_{\overline{5}}}{v_{\overline{45}}}\varepsilon\Big|\\
\nonumber&&\sin^2\theta_{12}\simeq\frac{1}{3}+\frac{1}{18}\Big[\frac{y^d_{12}}{y^d_{22}}\frac{v_{\overline{5}}}{v_{\overline{45}}}\varepsilon+(\frac{y^d_{12}}{y^d_{22}}\frac{v_{\overline{5}}}{v_{\overline{45}}}\varepsilon)^{*}\Big]\\
\label{39}&&\sin\theta^2_{23}\simeq\frac{1}{2}+\frac{1}{144}\Big|\frac{y^d_{12}}{y^d_{22}}\frac{v_{\overline{5}}}{v_{\overline{45}}}\varepsilon\Big|^2
\end{eqnarray}
Taking into account the results for quark mixing shown in
Eq.(\ref{21}), we have
$|V_{us}|\simeq|\frac{y^d_{21}}{2y^d_{22}}\frac{v_{\overline{5}}}{v_{\overline{45}}}\varepsilon|\sim\lambda_c$.
As a result, the model predicts the deviation of the lepton mixing
from the TB pattern as follows
\begin{eqnarray}
\nonumber&&\sin\theta_{13}\sim\frac{\lambda_c}{3\sqrt{2}}\simeq2.97^{\circ}\\
\nonumber&&|\sin^2\theta_{12}-\frac{1}{3}|\sim\frac{2}{9}\lambda_c\\
\label{40}&&|\sin^2\theta_{23}-\frac{1}{2}|\sim\frac{\lambda^2_c}{36}
\end{eqnarray}
The lepton mixing angles are predicted to be in agreement at
$3\sigma$ error with the experimental data
\cite{Strumia:2006db,Schwetz:2008er,Fogli:Indication}. It is remarkable that Eq.(\ref{40}) belongs to a set of  well-known leptonic mixing sum rules \cite{King:2005bj,Antusch:2007ib}, and the same results have been obtained in Ref.\cite{Hagedorn:2010th}.

\subsection{High dimensional Weinberg operators}
In the previous section, the neutrinos acquire masses via the
see-saw mechanism. It is interesting to note that the higher
dimensional Weinberg operator could also contribute to the neutrino
masses directly. In the present model, these effective light
neutrino mass operators are\footnote{Concretely the operator
$(FF)_{3_1}\chi H_{45}H_{45}$ denotes
$y_{abc}(F_a)_{\alpha}(F_b)_{\beta}\chi_c(H_{45})^{\gamma\alpha}_{\delta}(H_{45})^{\delta\beta}_{\gamma}$,
where the Greek indices are contracted in the $SU(5)$ space, and the
Latin indices are contracted in the $S_4$ space, the coefficient
with three $S_4$ indices $y_{abc}$ is the Clebsch-Gordon coefficient
of the $S_4$ group, its value can be read directly from the product
decomposition rules shown in Appendix A, so that the effective
operator is invariant under the flavor group $S_4$. The contractions
of the remaining operators in Eq.(\ref{41}) can be read out
similarly.}
\begin{equation}
\label{41}w^{eff}_{\nu}=\frac{y_{\nu1}}{\Lambda^2}(FF)_{3_1}\chi
H_5H_5+\frac{y_{\nu2}}{\Lambda^2}(FF)_{2}\varphi
H_5H_5+\frac{y_{\nu3}}{\Lambda^2}(FF)_{3_1}\chi
H_{45}H_{45}+\frac{y_{\nu4}}{\Lambda^2}(FF)_{2}\varphi H_{45}H_{45}
\end{equation}
Taking into account the vacuum alignment
$\langle\chi\rangle=v_{\chi}(1,1,1)^{T}$ and
$\langle\varphi\rangle=v_{\varphi}(1,1)^{T}$, the Weinberg operators
in $w^{eff}_{\nu}$ lead to the following effective light neutrino
mass matrix
\begin{equation}
\label{42}M^{eff}_{\nu}=\left(\begin{array}{ccc}2\alpha&-\alpha+\beta&-\alpha+\beta\\
-\alpha+\beta&2\alpha+\beta&-\alpha\\
-\alpha+\beta&-\alpha&2\alpha+\beta
\end{array}\right)
\end{equation}
where
\begin{eqnarray}
\nonumber&&\alpha=(2y_{\nu1}\frac{v^2_5}{\Lambda}+24y_{\nu3}\frac{v^2_{45}}{\Lambda})\frac{v_{\chi}}{\Lambda}\\
\label{43}&&\beta=(2y_{\nu2}\frac{v^2_5}{\Lambda}+24y_{\nu4}\frac{v^2_{45}}{\Lambda})\frac{v_{\varphi}}{\Lambda}
\end{eqnarray}
The mass matrix $M^{eff}_{\nu}$ is exactly diagonalized by the TB
mixing matrix
\begin{equation}
\label{44}U^{T}_{TB}M^{eff}_{\nu}U_{TB}={\rm
diag}(m^{eff}_1,m^{eff}_2,m^{eff}_3)
\end{equation}
As a result, the lepton mixing angles displayed in Eq.(\ref{39}) are
not corrected even if the Weinberg operators are taken into account.
The mass eigenvalues $m^{eff}_{1,2,3}$ are given by
\begin{eqnarray}
\nonumber&&m^{eff}_1=3\alpha-\beta\\
\nonumber&&m^{eff}_2=2\beta\\
\label{45}&&m^{eff}_3=3\alpha+\beta
\end{eqnarray}
Comparing with the see-saw mechanism induced masses in
Eq.(\ref{28}), we have
\begin{equation}
\label{46}\frac{m^{eff}_i}{m_i}\sim\frac{v^2_{\chi}}{\Lambda^2}
\end{equation}
It is obvious that the contributions of the Weinberg operator are
highly suppressed relative to those induced by the see-saw
mechanism, so that they can be completely negligible.

\section{Vacuum alignment}

\begin{table}
\begin{center}
\begin{tabular}{|c|c|c|c|c|c|}\hline\hline
Fields& $~~\chi^{0}~~$& $~~\varphi^{0}~~$  & $~~\phi^{0}~~$ & $~~\rho^{0}~~$&
$~~\Delta^{0}~~$
\\\hline

$\rm{S_4}$& $3_2$  & $2$& $3_1$ & $1_1$ &$3_1$
\\\hline

$\rm{Z_{3}}$& 1 &1 & $\omega^2$  & $\omega^2$   & $\omega$
\\\hline

$\rm{Z_{4}}$& 1 &1 &-1 & -1& -1   \\ \hline

$U(1)_R$& 2 &2 &2 &2& 2\\ \hline\hline
\end{tabular}
\end{center}
\caption{\label{tab:driving}Driving fields and their transformation
rules under the symmetry group $S_4\times Z_3\times Z_4$. }
\end{table}

In this section we discuss the minimization of the scalar potential
in order to justify the vacuum alignment used in the previous
section. As usual we introduce a global continuous $U(1)_R$ symmetry
which contain the discrete $R-$parity as a subgroup. The flavon and
GUT Higgs fields are uncharged under $U(1)_R$, the supermultiplets
containing the standard model matter fields and the adjoint field
$A$ carry $U(1)_R$ charge +1. Moreover, we include additional gauge
singlets, the so called driving fields $\chi^0$, $\varphi^0$,
$\phi^0$, $\rho^0$ and $\Delta^0$ with $U(1)_R$ charge +2. They transform in a
non-trivial way under the flavor symmetry $S_4\times Z_3\times Z_4$,
as is presented in Table \ref{tab:driving}. Since the driving fields
carry +2 unit $U(1)_R$ charge, they enter linearly into the
superpotential. The LO superpotential depending on the driving
fields, which is invariant under the flavor symmetry, is given by
\begin{eqnarray}
\nonumber
w_v&=&f_1\chi^0(\chi\varphi)_{3_2}+f_2\chi^0\chi\zeta+f_3\varphi^0(\chi\chi)_2+f_4\varphi^0(\varphi\varphi)_2+f_5\varphi^0\varphi\zeta
+g_1\phi^0(\phi\phi)_{3_1}\\
\label{47}&&+g_2\phi^0(\eta\phi)_{3_1}+g_3\rho^0(\phi\phi)_{1_1}+g_4\rho^0(\eta\eta)_{1_1}+h_1\Delta^0(\Delta\Delta)_{3_1}+h_2\Delta^0\Delta\xi
\end{eqnarray}
In the SUSY limit, the equations for the minimum of the scalar
potential are obtained by deriving $w_v$ with respect to each
component of the driving fields
\begin{eqnarray}
\nonumber&&\frac{\partial
w_v}{\partial\chi^0_1}=f_1(\varphi_1\chi_2-\varphi_2\chi_3)+f_2\chi_1\zeta=0\\
\nonumber&&\frac{\partial
w_v}{\partial\chi^0_2}=f_1(\varphi_1\chi_1-\varphi_2\chi_2)+f_2\chi_3\zeta=0\\
\nonumber&&\frac{\partial
w_v}{\partial\chi^0_3}=f_1(\varphi_1\chi_3-\varphi_2\chi_1)+f_2\chi_2\zeta=0\\
\nonumber&&\frac{\partial w_v}{\partial
\varphi^0_1}=f_3(\chi^2_3+2\chi_1\chi_2)+f_4\varphi^2_1+f_5\varphi_2\zeta=0\\
\label{48}&&\frac{\partial
w_v}{\partial\varphi^0_2}=f_3(\chi^2_2+2\chi_1\chi_3)+f_4\varphi^2_2-f_5\varphi_1\zeta=0
\end{eqnarray}
This set of equations are satisfied by three types of vacuum
alignment
\begin{equation}
\label{49}\langle\chi\rangle=v_{\chi}\left(\begin{array}{c}1\\
1\\
1\end{array}\right),~~~~~~\langle\varphi\rangle=v_{\varphi}\left(\begin{array}{c}1\\
1\end{array}\right),~~~~~~\langle\zeta\rangle=0
\end{equation}
with the conditions
\begin{equation}
\label{50}v^2_{\chi}=-\frac{f_4}{3f_3}v^2_{\varphi},~~~~~v_{\varphi}~
{\rm undetermined}
\end{equation}
The second is
\begin{equation}
\label{add3}\langle\chi\rangle=v_{\chi}\left(\begin{array}{c}1\\
1\\
1\end{array}\right),~~~~~~\langle\varphi\rangle=v_{\varphi}\left(\begin{array}{c}1\\
-1\end{array}\right),~~~~~~\langle\zeta\rangle=v_{\zeta}
\end{equation}
with the relations
\begin{equation}v^2_{\chi}=-\frac{f^2_2f_4+2f_1f_2f_5}{12f^2_1f_3}v^2_{\zeta},~~~v_{\varphi}=-\frac{f_2}{2f_1}v_{\zeta}
\end{equation}
where $v_{\zeta}$ is undetermined. The third solution is
\begin{equation}
\label{add4}\langle\chi\rangle=\left(\begin{array}{c}0\\0\\
0
\end{array}\right),~~~~~~\langle\varphi\rangle=v_{\varphi}\left(\begin{array}{c}1\\
-1\end{array}\right),~~~~~~\langle\zeta\rangle=v_{\zeta}
\end{equation}
with
\begin{equation}
\label{add5}v_{\varphi}=\frac{f_5}{f_4}v_{\zeta},~~~~~v_{\zeta}~
{\rm undetermined}
\end{equation}
Thus, without assuming any fine-tuning among the parameters $f_i(1=1-5)$, the VEVs $v_{\chi}$ and $v_{\varphi}$ are expected to be of the same order of magnitude for the three cases,
\begin{equation}
\label{add2}v_{\chi}\sim v_{\varphi}
\end{equation}
Only the first alignment can produce the results in the previous sections, we need
of some soft masses in order to discriminate it as the lowest minimum
of the scalar potential, since the values of the scalar potential for the three solutions are exactly the same in the SUSY limit. It is well-known that the soft mass usually is of order TeV, consequently the difference of the scalar potential for different vacuum solutions is marginal comparing with the flavon VEVs. As has been shown in the previous section,
at LO the $S_4$ flavor symmetry is broken by the VEV of $\chi$ and
$\varphi$ in the neutrino sector. The flavon fields $\phi$, $\eta$,
$\Delta$ and $\xi$ are involved in generating the quark and charged
lepton masses, their vacuum configurations are determined by
\begin{eqnarray}
\nonumber&&\frac{\partial
w_v}{\partial\phi^0_1}=2g_1(\phi^2_1-\phi_2\phi_3)+g_2(\eta_1\phi_2+\eta_2\phi_3)=0\\
\nonumber&&\frac{\partial
w_v}{\partial\phi^0_2}=2g_1(\phi^2_2-\phi_1\phi_3)+g_2(\eta_1\phi_1+\eta_2\phi_2)=0\\
\nonumber&&\frac{\partial
w_v}{\partial\phi^0_3}=2g_1(\phi^2_3-\phi_1\phi_2)+g_2(\eta_1\phi_3+\eta_2\phi_1)=0\\
\label{51}&&\frac{\partial w_v}{\partial\rho^0}=g_3(\phi^2_1+2\phi_2\phi_3)+2g_4\eta_1\eta_2=0
\end{eqnarray}
\begin{eqnarray}
\nonumber&&\frac{\partial w_v}{\partial
\Delta^0_1}=2h_1(\Delta^2_1-\Delta_2\Delta_3)+h_2\Delta_1\xi=0\\
\nonumber&&\frac{\partial w_v}{\partial
\Delta^0_2}=2h_1(\Delta^2_2-\Delta_1\Delta_3)+h_2\Delta_3\xi=0\\
\label{52}&&\frac{\partial w_v}{\partial
\Delta^0_3}=2h_1(\Delta^2_3-\Delta_1\Delta_2)+h_2\Delta_2\xi=0
\end{eqnarray}
The equations in Eq.(\ref{51}) lead to two different vacuum configurations, the first is
\begin{equation}
\label{53}\langle\phi\rangle=\left(\begin{array}{c}0\\
v_{\phi}\\0\end{array}\right),~~~~~~\langle\eta\rangle=\left(\begin{array}{c}0\\
v_{\eta}\end{array}\right)
\end{equation}
with
\begin{equation}
\label{54}v_{\phi}=-\frac{g_2}{2g_1}v_{\eta},~~~v_{\eta}~{\rm
undetermined}
\end{equation}
The second is
\begin{equation}
\label{57}\langle\phi\rangle=v_{\phi}\left(\begin{array}{c}1\\
1\\
1\end{array}\right),~~~~~\langle\eta\rangle=v_{\eta}\left(\begin{array}{c}1\\-1\end{array}\right)
\end{equation}
with $v^2_{\phi}=\frac{2g_4}{3g_3}v^2_{\eta}$ and $v_{\eta}$ undetermined. Obviously $v_{\phi}$ and $v_{\eta}$ are expected to be of the same order for both solutions. As before, we select the first vacuum configuration. The equations in Eq.(\ref{52}) admit three un-equivalent solutions, the first is
\begin{equation}
\label{58}\langle\Delta\rangle=\left(\begin{array}{c}v_{\Delta}\\0\\0\end{array}\right),~~~~~\langle\xi\rangle=v_{\xi}
\end{equation}
with $v_{\Delta}=-\frac{h_2}{2h_1}v_{\xi}$ and $v_{\xi}$
undetermined. In a similar way, the VEVs $v_{\Delta}$ and $v_{\xi}$ should be of the same order. The second solution is
\begin{equation}
\label{59}\langle\Delta\rangle=v_{\Delta}\left(\begin{array}{c}1\\1\\1\end{array}\right),~~~~~~\langle\xi\rangle=0
\end{equation}
with $v_{\Delta}$ undetermined. The third vacuum configuration is
\begin{equation}
\label{add5}\langle\Delta\rangle=\left(\begin{array}{c}0\\0\\0\end{array}\right),~~~~~~\langle\xi\rangle=v_{\xi}
\end{equation}
and $v_{\xi}$ remaining undetermined. The first vacuum solution is chosen in the present model. Acting on the vacuum configurations of Eq.(\ref{49}), Eq.(\ref{53}) and Eq.(\ref{58}) with the elements of the flavor symmetry group $S_4$, we can generate other minima of the scalar potential. These new minima are physically equivalent to the original set,
and they all lead to the same physics, i.e. fermion masses and flavor mixings. As a result, without loss of generality we can analyze the model by choosing exactly the vacuum in Eqs.(\ref{49},\ref{53},\ref{58}) as local minimum, and the different scenarios are related by field redefinitions. Since no superpotential couplings of positive mass dimension are involved in the flavon superpotential, the trivial solution with all flavon VEVs vanishing can not be excluded. However, by taking into account the contribution of the soft mass terms, we can discriminate the configuration in Eqs.(\ref{49},\ref{53},\ref{58}) as the lowest minimum of the scalar potential\footnote{We consider the soft terms involving $\chi$ and $\varphi$, which is generally written as $m^2_{\chi}|\chi|^2+m^2_{\varphi}|\varphi|^2+\tilde{m}^2_{\chi}\chi^2+\tilde{m}^2_{\varphi}\varphi^2$. By choosing $m^2_{\chi}$, $m^2_{\varphi}$, $\tilde{m}^2_{\chi}$ and $\tilde{m}^2_{\varphi}<0$, the vacuum shown in Eqs.(\ref{49},\ref{53},\ref{58}) are more stable than the vanishing VEVs configuration.}. Regarding the size of the flavon VEVs, we have the relations $v_{\chi}\sim v_{\varphi}$, $v_{\phi}\sim v_{\eta}$ and $v_{\Delta}\sim v_{\xi}$, as have been demonstrated above \footnote{In the absence of specific dynamical tricks (see Ref.\cite{Lin:2009bw} for a model in which such a trick is implemented), the uncorrelated VEVs naturally have the same order of magnitude, this is consistent with the the constraints from the measured mass hierarchies and flavor mixing. Moreover, we note that the correlation of scales of more flavon VEVs can be achieved by adding further driving fields, whereas this procedure would introduce more fields and free parameters.}. Furthermore, The magnitudes of the flavon VEVs are determined by the patterns of fermion mass hierarchy and mixing. From the LO predictions presented in the previous sections, we find that in order to reproduce the correct patterns of fermion masses and
flavor mixings, a common order of magnitude for the VEVs
scaled by the cutoff $\Lambda$ is expected
\begin{equation}
\label{60}\frac{v_{\chi}}{\Lambda}\sim\frac{v_{\varphi}}{\Lambda}\sim\frac{v_{\phi}}{\Lambda}\sim\frac{v_{\eta}}{\Lambda}\sim\frac{v_{\Delta}}{\Lambda}\sim\frac{v_{\xi}}{\Lambda}\sim\lambda^2_c
\end{equation}
Moreover, we will show in the following that the successful LO
predictions are not destroyed by the subleading corrections when the
vacuum alignment is chosen as has been stated above. Similar to the fact that the F-terms of the driving fields are the origin of the alignment of the flavon VEVs, we can derive the vacuum
structure of the driving fields from the F-terms of the flavon fields. As all terms in the flavon superpotential are linear in the driving fields,
the configuration in which all these fields have vanishing VEVs is in any case a solution.
Moreover, we have checked that this is unique vacuum configuration of driving fields in our model.

\section{Subleading corrections}
At the next level of expansion in $1/\Lambda$, the superpotentials
$w_{\nu}$, $w_u$, $w_d$ and $w_v$ are corrected by higher
dimensional operators whose contributions are suppressed by at least
one power of $1/\Lambda$. The corrections to $w_v$ result in small
deviations from the LO vacuum alignment thus affect the results for
fermion masses and mixings. In addition the fermion mass and mixing
matrices are corrected by the subleading operators in $w_{\nu}$,
$w_u$ and $w_d$. In the following we shall first present the
analysis for the subleading corrections to the vacuum alignment,
then move to the corrections to fermion mass matrices.
\subsection{Corrections to the vacuum alignment}
We detail the discussion of this issue in Appendix B, here we only
present the results. The vacuum configuration is shifted into
\begin{eqnarray}
\nonumber&&\langle\chi\rangle=\left(\begin{array}{c}v_{\chi}+\delta
v_{\chi1}\\v_{\chi}+\delta v_{\chi2}\\v_{\chi}+\delta v_{\chi3}
\end{array}\right),~~~~~\langle\varphi\rangle=\left(\begin{array}{c}v_{\varphi}\\v_{\varphi}+\delta
v_{\varphi2}\end{array}\right),~~~~~\langle\zeta\rangle=\delta v_{\zeta}\\
\nonumber&&\langle\phi\rangle=\left(\begin{array}{c}\delta
v_{\phi1}\\v_{\phi}+\delta v_{\phi2}\\ \delta
v_{\phi3}\end{array}\right),~~~~~\langle\eta\rangle=\left(\begin{array}{c}\delta
v_{\eta1}\\v_{\eta}\end{array}\right)\\
\label{61}&&\langle\Delta\rangle=\left(\begin{array}{c}v_{\Delta}+\delta
v_{\Delta1}\\ \delta v_{\Delta2}\\ \delta
v_{\Delta3}\end{array}\right),~~~~~\langle\xi\rangle=v_{\xi}
\end{eqnarray}
where $v_{\varphi}$, $v_{\eta}$ and $v_{\xi}$ remain undetermined.
Since we typically have $VEV/\Lambda\sim\lambda^2_c$ at LO, we
expect the order of magnitude of the shifts as follows
\begin{eqnarray}
\nonumber&&\delta v_{\chi1}/v_{\chi}\sim\lambda^2_c,~~~\delta
v_{\chi2}/v_{\chi}\sim\lambda^2_c,~~~\delta
v_{\chi3}/v_{\chi}\sim\lambda^2_c,~~~\delta
v_{\varphi2}/v_{\varphi}\sim\lambda^2_c,~~~\delta v_{\zeta}/v_{\chi}\sim\lambda^2_c\\
\nonumber&&\delta v_{\phi1}/v_{\phi}\sim\lambda^4_c,~~~\delta
v_{\phi2}/v_{\phi}\sim\lambda^4_c,~~~\delta
v_{\phi3}/v_{\phi}\sim\lambda^4_c,~~~\delta
v_{\eta1}/v_{\eta}\sim\lambda^4_c\\
\label{62}&&\delta v_{\Delta1}/v_{\Delta}\sim\lambda^4_c,~~~\delta
v_{\Delta2}/v_{\Delta}\sim\lambda^4_c,~~~\delta
v_{\Delta3}/v_{\Delta}\sim\lambda^4_c
\end{eqnarray}
From Appendix B, we can see that the subleading operators linear in $\chi^0$ and $\varphi^0$ are suppressed by $\Lambda$, while the subleading operators linear in $\phi^0$, $\rho^0$ and $\Delta^0$ are suppressed by $\Lambda^2$, due to the constraint of the flavor symmetry $S_4\times Z_3\times Z_4$. Consequently the shifts are of order $\lambda^2_c$ or $\lambda^4_c$ with respect to the LO VEVs, as is displayed in Eq.(\ref{62}).

\subsection{Corrections to fermion masses and mixings}
The fermion masses and mixings receive corrections from both the
shifted VEVs and the higher order terms in the superpotentials
$w_{\nu}$, $w_u$ and $w_d$. We can obtain the mass matrix by
inserting the modified VEV into the LO operators plus the
contribution of the higher dimensional operators evaluated with the
LO VEVs. For convenience, we denote $\chi$, $\varphi$ and $\zeta$ with
$\Phi_{\nu}$, $\Delta$ and $\xi$ with $\Phi_{f1}$, $\phi$ and $\eta$
with $\Phi_{f2}$ in the following.
\subsubsection{Corrections to up quark sector}
First we discuss the corrections to the up type quark mass matrix
coming from the modified vacuum alignment. Plugging the shifted
vacuum configuration shown in Eq.(\ref{61}) into the LO
superpotential in Eq.(\ref{3}), we find that the (12) element
receives a correction of order $\varepsilon^4$ from the operators
$(TT)_{1_1}(\phi\phi)_{1_1}H_5$ and $(TT)_{1_1}(\eta\eta)_{1_1}H_5$.
The operator $(TT)_2(\phi\phi)_2H_5$ induces corrections of order
$\varepsilon^4$ to both (11) and (22) elements. The corrections to
the (13) element due to modified VEVs are of order $\varepsilon^3$,
they arise from the contractions $TT_3(\phi\chi)_2H_5$,
$TT_3(\eta\varphi)_2H_5$ and $TT_3\eta H_{45}$. Meanwhile,
$TT_3(\phi\chi)_2H_5$ introduces correction of order $\varepsilon^3$
to the (23) element.

Then we come to discuss the corrections caused by the higher
dimensional operators in the matter superpotential $w_u$. We note
that all corrections to the (33) element can be absorbed into the
coupling $y_t$ of the LO operator $T_3T_3H_5$, and any operator
comprising the superfields $T_3T_3H_{45}$ ($TTH_{45}$) and arbitrary
number of flavon fields gives a vanishing contribution to the up
quark mass matrix. Due to the auxiliary symmetry $Z_3\times Z_4$,
the subleading corrections to (11), (12), (21) and (22) elements
arise at order $1/\Lambda^4$, they come from the following
contraction
\begin{equation}
\label{63}\frac{1}{\Lambda^4}TT\Phi_{\nu}\Phi_{\nu}\Phi_{f2}\Phi_{f2}H_5
\end{equation}
Consequently the corrections to the ($ij$)($i,j=1,2$) element from
the high dimensional operators are of the same order as those from
the shifted vacuum. In the same way, we find the (13) and (23)
elements are corrected by the following contractions
\begin{equation}
\label{64}\frac{1}{\Lambda^3}TT_3\Phi_{f1}\Phi_{f2}\Phi_{f2}H_5,~~~~~\frac{1}{\Lambda^3}TT_3\Phi_{\nu}\Phi_{\nu}\Phi_{f2}H_{45}
\end{equation}
Substituting the LO VEVs into the above contractions, we notice that
the corrections to the (23) and (32) elements originate from the
latter contraction, whereas both operators contribute to (13) and
(31) elements. In short summary, the up type quark mass matrix are
corrected by both the deviations from the LO VEV alignment and the
higher dimensional operators allowed by the flavor symmetry. We can
parameterize the up quark mass matrix as
\begin{equation}
\label{65}M_u=\left(\begin{array}{ccc}
8y^u_{11}\varepsilon^4v_5&8y^u_{12}\varepsilon^4v_5&4y^u_{13}\varepsilon^2v_5+8y^{u'}_{13}\varepsilon^3v_{45}\\
8y^u_{12}\varepsilon^4v_5&8y^u_{22}\varepsilon^2v_5&8y^u_{23}\varepsilon
v_{45}+4y^{u'}_{23}\varepsilon^2v_5\\
4y^u_{13}\varepsilon^2v_5-8y^{u'}_{13}\varepsilon^3v_{45}&-8y^u_{23}\varepsilon
v_{45}+4y^{u'}_{23}\varepsilon^2v_5&8y^u_{33}v_5
\end{array}\right)
\end{equation}
where $y^u_{ij}$ ($i,j=1,2,3$) are complex numbers with absolute
value of order one, and they are linear combinations of the leading
and subleading coefficients. We note that $y^u_{33}$ coincides with
the LO parameter $y_t$ up to higher order corrections which are due
to two flavons and three flavons insertions in the operator
$T_3T_3H_5$. Similarly the parameters $y^u_{13}$, $y^u_{22}$,
$y^u_{23}$ and $y^{u'}_{23}$ are determined by the LO couplings in
Eq.(\ref{3}) up to small corrections of relative order $\varepsilon$
or $\varepsilon^2$. The mass matrix $M_u$ in Eq.(\ref{65}) leads to
the up type quark masses
\begin{eqnarray}
\nonumber&&m_u\simeq\Big|\Big[8y^u_{11}v_5-2\frac{(y^u_{13})^2}{y^u_{33}}v_5+\frac{2(y^u_{13})^2(y^u_{23})^2v_5v^2_{45}}{y^u_{22}(y^u_{33})^2v^2_5+y^u_{33}(y^u_{23})^2v^2_{45}}\Big]\varepsilon^4\Big|\\
\nonumber&&m_c\simeq\Big|8\Big[y^u_{22}v_5+\frac{(y^u_{23})^2}{y^u_{33}}\frac{v^2_{45}}{v_5}\Big]\varepsilon^2\Big|\\
\label{66}&&m_t\simeq|8y^u_{33}v_5|
\end{eqnarray}
The correct hierarchies among the up quark masses are obtained.
\subsubsection{Corrections to down quark and charged lepton sector}
Plugging the shifted vacuum of $\phi$ into the LO operator $T_3F\phi
H_{\overline{5}}$ leads to corrections to the (13), (23) and (33)
elements of $M_d$ of order $\varepsilon^3v_{\overline{5}}$, this
amounts to a rescaling of the parameters $y^d_{13}$, $y^d_{23}$ and
$y^d_{33}$. If the non-zero shifts $\delta v_{\Delta1,2,3}$ are
taken into account, the LO operators $TF\Delta\Delta
H_{\overline{45}}$ and $TF\Delta\xi H_{\overline{45}}$ introduces
corrections of order $\varepsilon^4v_{\overline{45}}$ to the (22),
(31), (32), (11), (12) and (21) elements which also receive
corrections of order $\varepsilon^4v_{\overline{5}}$ from the
operators $TF{\cal O}^{(5)}H_{\overline{5}}$. At LO the
superpotential $w_d$ is expanded to $1/\Lambda^3$, it is corrected
by the following subleading operators
\begin{eqnarray}
\nonumber&&\frac{1}{\Lambda^4}T_3F\Phi_{\nu}\Phi_{f1}\Phi_{f2}\Phi_{f2}H_{\overline{5}},~~~\frac{1}{\Lambda^4}TF\Phi_{f1}\Phi_{f2}\Phi_{f2}\Phi_{f2}H_{\overline{5}},~~~\frac{1}{\Lambda^4}TF\Phi_{f1}\Phi_{f1}\Phi_{f1}\Phi_{f1}H_{\overline{5}},\\
\label{67}&&\frac{1}{\Lambda^4}TF\Phi_{\nu}\Phi_{\nu}\Phi_{f1}\Phi_{f1}H_{\overline{45}}
\end{eqnarray}
With the LO VEVs, the above high dimensional operators lead to
corrections of order $\varepsilon^4v_{\overline{5}}$ or
$\varepsilon^4v_{\overline{45}}$ in each entry of $M_d$. Note that
the subleading terms involving the fields combination
$T_3FH_{\overline{45}}$ arise at order $1/\Lambda^5$ with the
insertion of five flavon fields. Therefore we conclude that the NLO
corrections to the down quark and charged lepton mass matrices can be reabsorbed into a
redefinition of the LO parameters whose order of magnitudes are not
changed. As a result, we can parameterize the down quark and charged lepton
mass matrices in the same way as Eq.(\ref{11}) and Eq.(\ref{12}), and the parameters $y^d_{ij}(i,j=1,2,3)$ are still used to avoid introducing extra unnecessary parameters. However, we should keep in mind that the value of $y^d_{ij}$ is different from the corresponding LO one due to the subleading corrections of relative order $\varepsilon$ or $\varepsilon^2$. Since the down quark and charged lepton
mass matrices remain the same form, the down quark and charged lepton masses are still
given by Eq.(\ref{14}) and Eq.(\ref{15}) respectively.
Starting from the quark mass matrices $M_d$ in Eq.(\ref{11}) and $M_u$ in
Eq.(\ref{65}), we can straightforwardly find the CKM mixing matrix
as follows
\begin{eqnarray}
\nonumber&&V_{ud}\simeq V_{cs}\simeq V_{tb}\simeq1\\
\nonumber&&V^{*}_{us}\simeq-V_{cd}\simeq\frac{y^d_{21}}{2y^d_{22}}\frac{v_{\overline{5}}}{v_{\overline{45}}}\varepsilon+\frac{1}{2}\frac{y^u_{13}y^u_{23}v_5v_{45}}{y^u_{22}y^u_{33}v^2_5+(y^u_{23})^2v^2_{45}}\varepsilon\\
\nonumber&&V^{*}_{ub}\simeq2\frac{y^d_{22}}{y^d_{33}}\frac{v_{\overline{45}}}{v_{\overline{5}}}\varepsilon+\Big[\frac{y^{d'}_{31}}{y^d_{33}}-\frac{y^u_{13}}{2y^u_{33}}+\frac{1}{2}\frac{y^u_{13}(y^u_{23})^2v^2_{45}}{y^u_{22}(y^u_{33})^2v^2_5+y^u_{33}(y^u_{23})^2v^2_{45}}\Big]\varepsilon^2\\
\nonumber&&V^{*}_{cb}\simeq-V_{ts}\simeq\frac{y^u_{23}}{y^u_{33}}\frac{v_{45}}{v_5}\varepsilon\\
\label{68}&&V_{td}\simeq-2\frac{y^d_{22}}{y^d_{33}}\frac{v_{\overline{45}}}{v_{\overline{5}}}\varepsilon+\Big[\frac{y^d_{21}y^u_{23}}{2y^d_{22}y^u_{33}}\frac{v_{45}}{v_5}\frac{v_{\overline{5}}}{v_{\overline{45}}}-\frac{y^{d'}_{31}}{y^d_{33}}+\frac{y^u_{13}}{2y^u_{33}}\Big]\varepsilon^2
\end{eqnarray}
We see that the CKM parameters are determined by the LO results up to small corrections, which is absorbed in the redefinition of the parameters.
The successful LO predictions for the order of
magnitudes of the CKM matrix elements are not spoiled by the
subleading corrections.
\subsubsection{Corrections to neutrino sector}
The superpotential $w_{\nu}$ in Eq.(\ref{22}) is corrected by terms
with more flavons insertion. 
The VEV shifts in the LO operator do not affect the Dirac mass, the
NLO corrections to the neutrino Dirac couplings are
\begin{equation}
\label{69}\frac{y_{\nu1}}{\Lambda}(FA)_{3_1}\chi
H_{45}+\frac{y_{\nu2}}{\Lambda}(FA)_2\varphi H_{45}
\end{equation}
Using the alignment of $\chi$ and $\varphi$ given in Eq.(\ref{49}),
the NLO corrections to the Dirac mass matrices read
\begin{eqnarray}
\nonumber&&\delta
M^{D}_{\rho_3}=-\frac{3}{2}v_{45}\left(\begin{array}{ccc}2y_{\nu1}\frac{v_{\chi}}{\Lambda}&-y_{\nu1}\frac{v_{\chi}}{\Lambda}+y_{\nu2}\frac{v_{\varphi}}{\Lambda}&-y_{\nu1}\frac{v_{\chi}}{\Lambda}+y_{\nu2}\frac{v_{\varphi}}{\Lambda}\\
-y_{\nu1}\frac{v_{\chi}}{\Lambda}+y_{\nu2}\frac{v_{\varphi}}{\Lambda}&2y_{\nu1}\frac{v_{\chi}}{\Lambda}+y_{\nu2}\frac{v_{\varphi}}{\Lambda}&-y_{\nu1}\frac{v_{\chi}}{\Lambda}\\
-y_{\nu1}\frac{v_{\chi}}{\Lambda}+y_{\nu2}\frac{v_{\varphi}}{\Lambda}&-y_{\nu1}\frac{v_{\chi}}{\Lambda}&2y_{\nu1}\frac{v_{\chi}}{\Lambda}+y_{\nu2}\frac{v_{\varphi}}{\Lambda}
\end{array}\right)\\
\label{70}&&\delta M^{D}_{\rho_0}=-\frac{\sqrt{15}}{3}\delta
M^{D}_{\rho_3}
\end{eqnarray}
We note that the subleading corrections $\delta M^{D}_{\rho_3}$ and
$\delta M^{D}_{\rho_0}$ are still compatible with the TB mixing. The
Majorana mass matrices of $\rho_3$ and $\rho_0$ are modified by the
terms
\begin{eqnarray}
\nonumber&&\lambda_1A^2\delta\chi+\lambda_2A^2\delta\varphi+\frac{x_A}{\Lambda}A^2\chi
H_{24}+\frac{x_B}{\Lambda}A^2\varphi
H_{24}+\frac{x_{C1}}{\Lambda}(AA)_{1_1}(\phi\Delta)_{1_1}+\frac{x_{C2}}{\Lambda}(AA)_2(\phi\Delta)_2\\
\label{71}&&+\frac{x_{C3}}{\Lambda}(AA)_{3_1}(\phi\Delta)_{3_1}+\frac{x_D}{\Lambda}(AA)_{3_1}\phi\xi+\frac{x_E}{\Lambda}(AA)_{3_1}(\eta\Delta)_{3_1}+\frac{x_F}{\Lambda}(AA)_2\eta\xi
\end{eqnarray}
where $\delta\chi$ and $\delta\varphi$ denote the shifted vacua of
the flavons $\chi$ and $\varphi$. The operators $A^2\chi H_{24}$ and
$A^2\varphi H_{24}$ lead to mass splitting between the triplet
$\rho_3$ and the singlet $\rho_0$, and the resulting contributions
to the Majorana mass matrices have the same structure as the LO
predictions. Taking into account the possibility of absorbing the
corrections partly into the LO results, the remaining operators give
rise to three independent additional contributions to the Majorana
mass matrices. As a result, the next to leading order corrections to
$M^{M}_{\rho_3}$ and $M^{M}_{\rho_0}$ can be parameterized as
\begin{eqnarray}
\nonumber\delta
M^M_{\rho_3}&=&\left(\begin{array}{ccc}-6x_A\epsilon v_{\chi}&3x_A\epsilon v_{\chi}-3x_B\epsilon v_{\varphi}&3x_A\epsilon v_{\chi}-3x_B\epsilon v_{\varphi}\\
3x_A\epsilon v_{\chi}-3x_B\epsilon v_{\varphi}&-6x_A\epsilon v_{\chi}-3x_B\epsilon v_{\varphi}&3x_A\epsilon v_{\chi}\\
3x_A\epsilon v_{\chi}-3x_B\epsilon v_{\varphi}&3x_A\epsilon
v_{\chi}&-6x_A\epsilon v_{\chi}-3x_B\epsilon v_{\varphi}
\end{array}\right)\\
\label{72}&&+\left(\begin{array}{ccc}0&-\tilde{x}_C&\tilde{x}_D-\tilde{x}_E\\
-\tilde{x}_C&\tilde{x}_{D}+2\tilde{x}_{E}&0\\
\tilde{x}_D-\tilde{x}_E&0&2\tilde{x}_C
\end{array}\right)\frac{v^2_{\chi}}{\Lambda}\\
\nonumber\delta
M^M_{\rho_0}&=&\left(\begin{array}{ccc}-2x_A\epsilon v_{\chi}&x_A\epsilon v_{\chi}-x_B\epsilon v_{\varphi}&x_A\epsilon v_{\chi}-x_B\epsilon v_{\varphi}\\
x_A\epsilon v_{\chi}-x_B\epsilon v_{\varphi}&-2x_A\epsilon v_{\chi}-x_B\epsilon v_{\varphi}&x_A\epsilon v_{\chi}\\
x_A\epsilon v_{\chi}-x_B\epsilon v_{\varphi}&x_A\epsilon
v_{\chi}&-2x_A\epsilon v_{\chi}-x_B\epsilon v_{\varphi}
\end{array}\right)\\
\label{73}&&+\left(\begin{array}{ccc}0&-\tilde{x}_C&\tilde{x}_D-\tilde{x}_E\\
-\tilde{x}_C&\tilde{x}_{D}+2\tilde{x}_{E}&0\\
\tilde{x}_D-\tilde{x}_E&0&2\tilde{x}_C
\end{array}\right)\frac{v^2_{\chi}}{\Lambda}
\end{eqnarray}
where
$\epsilon=\frac{1}{\sqrt{30}}\frac{v_{24}}{\Lambda}$\footnote{The
$SU(5)$ GUT symmetry is broken down to the standard model symmetry by the VEV
of $H_{24}$, i.e. $\langle H_{24}\rangle=v_{24}/\sqrt{30}~{\rm
diag}(2,2,2,-3,-3)$.}. The first term in both Eq.(\ref{72}) and
Eq.(\ref{73}) represents the contribution of
$\frac{x_A}{\Lambda}A^2\chi H_{24}$ and
$\frac{x_B}{\Lambda}A^2\varphi H_{24}$, the second term denotes the
effects of the modified vacuum configuration and the subleading
terms with the form $AA\Phi_{f1}\Phi_{f2}$, which leads to
deviations from the TB mixing pattern in the neutrino sector. With
the LO contributions shown in Eqs.(\ref{23},\ref{24}) and the NLO
corrections in Eqs.(\ref{70},\ref{72},\ref{73}), the light neutrino
mass matrix can be obtained straightforwardly via the see-saw
formula. To first order in small parameters ${v_{\chi}}/{\Lambda}$,
${v_{\varphi}}/{\Lambda}$ and $\epsilon$, we find that the light
neutrino masses are given by
\begin{eqnarray}
\nonumber&&m_1\simeq\Big|\frac{2y^2_{\nu}v^2_5}{5(-3\lambda_1v_{\chi}+\lambda_2v_{\varphi})}+\frac{9(-3x_Av_{\chi}+x_Bv_{\varphi})y^2_{\nu}v^2_5}{10(-3\lambda_1v_{\chi}+\lambda_2v_{\varphi})^2}\epsilon+\frac{(2\tilde{x}_C-\tilde{x}_D+2\tilde{x}_E)v_{\chi}^2y^2_{\nu}v^2_5}{5(-3\lambda_1v_{\chi}+\lambda_2v_{\varphi})^2\Lambda}\Big|\\
\nonumber&&m_2\simeq\Big|-\frac{y^2_{\nu}v^2_5}{5\lambda_2v_{\varphi}}-\frac{9x_By^2_{\nu}v^2_5}{20\lambda^2_2v_{\varphi}}\epsilon+\frac{\tilde{x}_Dv^2_{\chi}y^2_{\nu}v^2_5}{10\lambda^2_2v^2_{\varphi}\Lambda}\Big|\\
\label{74}&&m_3\simeq\Big|-\frac{2y^2_{\nu}v^2_5}{5(3\lambda_1v_{\chi}+\lambda_2v_{\varphi})}-\frac{9(3x_Av_{\chi}+x_Bv_{\varphi})y^2_{\nu}v^2_5}{10(3\lambda_1v_{\chi}+\lambda_2v_{\varphi})^2}\epsilon+\frac{(2\tilde{x}_C+\tilde{x}_D+2\tilde{x}_E)v^2_{\chi}y^2_{\nu}v^2_5}{5(3\lambda_1v_{\chi}+\lambda_2v_{\varphi})^2\Lambda}\Big|
\end{eqnarray}
The lepton mixing angles are modified as
\begin{eqnarray}
\nonumber&\sin\theta_{13}&=|U_{e3}|\simeq\Big|\big(\frac{y^d_{12}}{6\sqrt{2}y^d_{22}}\frac{v_{\overline{5}}}{v_{\overline{45}}}\varepsilon\big)^{*}-\frac{\tilde{x}_Dv^2_{\chi}(3\lambda^{*}_1v^{*}_{\chi}+\lambda^*_2v^{*}_{\varphi})+(3\lambda_1v_{\chi}-\lambda_2v_{\varphi})\tilde{x}^{*}_D(v^*_{\chi})^2}{\sqrt{2}\,[|-3\lambda_1v_{\chi}+\lambda_2v_{\varphi}|^2-|3\lambda_1v_{\chi}+\lambda_2v_{\varphi}|^2]\Lambda}\\
\nonumber&&-\frac{(\tilde{x}_C-\tilde{x}_E)v^2_{\chi}(3\lambda^*_1v^*_{\chi}+\lambda^*_2v^*_{\varphi})+2\lambda_2v_{\varphi}(\tilde{x}^{*}_C-\tilde{x}^{*}_E)(v^{*}_{\chi})^2}{\sqrt{2}\,[4|\lambda_2v_{\varphi}|^2-|3\lambda_1v_{\chi}+\lambda_2v_{\varphi}|^2]\Lambda}\Big|\\
\nonumber&\sin^2\theta_{12}&\simeq\frac{1}{3}+\frac{1}{18}[\frac{y^d_{12}}{y^d_{22}}\frac{v_{\overline{5}}}{v_{\overline{45}}}\varepsilon+(\frac{y^d_{12}}{y^d_{22}}\frac{v_{\overline{5}}}{v_{\overline{45}}}\varepsilon)^{*}]+\frac{1}{3[|-3\lambda_1v_{\chi}+\lambda_2v_{\varphi}|^2-4|\lambda_2v_{\varphi}|^2]\Lambda}\\
\nonumber&&\times[(\tilde{x}_C+\tilde{x}_E)v^2_{\chi}(3\lambda^*_1v^{*}_{\chi}+\lambda^*_2v^{*}_{\varphi})+(\tilde{x}^*_C+\tilde{x}^*_E)(v^*_{\chi})^2(3\lambda_1v_{\chi}+\lambda_2v_{\varphi})]\\
\nonumber&\sin^2\theta_{23}&\simeq\frac{1}{2}-\frac{3\tilde{x}_Dv_{\chi}\lambda^*_1|v_{\chi}|^2+3\tilde{x}^{*}_Dv^*_{\chi}\lambda_1|v_{\chi}|^2}{2[|3\lambda_1v_{\chi}+\lambda_2v_{\varphi}|^2-|-3\lambda_1v_{\chi}+\lambda_2v_{\varphi}|^2]\Lambda}-\frac{3}{2[4|\lambda_2v_{\varphi}|^2-|3\lambda_1v_{\chi}+\lambda_2v_{\varphi}|^2]\Lambda}\\
\label{75}&&\times[(\tilde{x}_C-\tilde{x}_E)v^2_{\chi}(\lambda^*_1v^*_{\chi}+\lambda^*_2v^*_{\varphi})+(\tilde{x}^*_C-\tilde{x}^{*}_E)(v^*_{\chi})^2(\lambda_1v_{\chi}+\lambda_2v_{\varphi})]
\end{eqnarray}
We note that the subleading terms in the neutrino sector induce
corrections of order $\varepsilon$ to all three mixing angles.
and the deviation of $\theta_{23}$ from its TB value is mainly
determined by the NLO contributions in the neutrino sector. The
lepton mixing angles are still compatible with the current
experimental data. In particular, the reactor angle $\theta_{13}$ is
within the reach of next generation neutrino oscillation
experiments.
\section{Phenomenological consequences}
In this section, we shall present the predictions for some
phenomenologically interesting observables in our model. Here we are
particularly interested in the neutrino sector, both the LO and the
NLO contributions are taken into account in the following.

\subsection{Lepton mixing angles}

\begin{figure}
\begin{center}
\begin{tabular}{cc}
\includegraphics[scale=1.1,width=6.5cm]{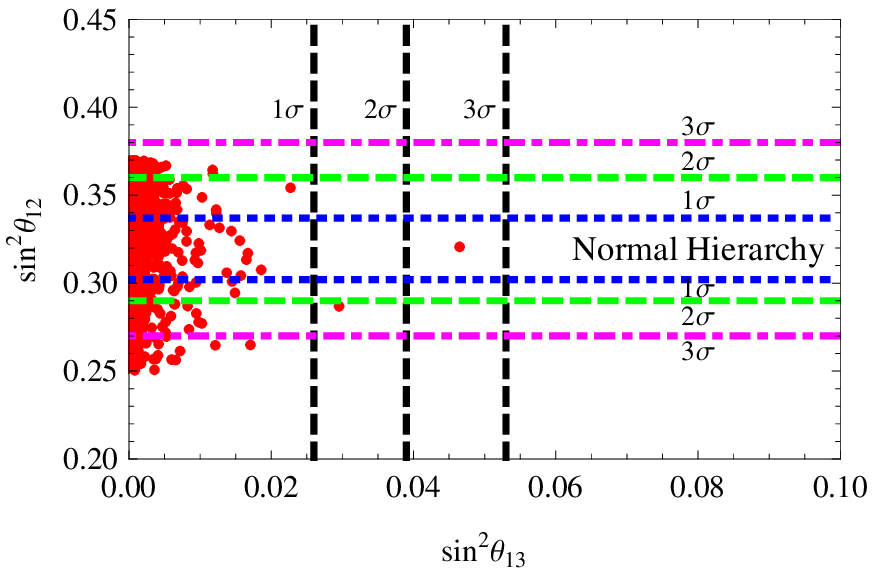}&\hspace{0.5cm}\includegraphics[scale=1.1,width=6.5cm]{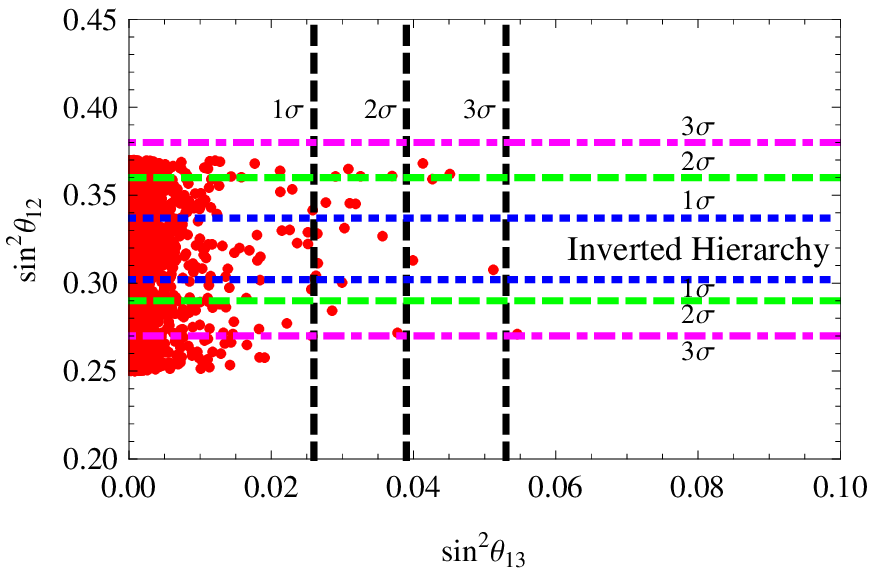}\\
~~~~~~(a)&~~~~~~~~~~(b)\\\\
\includegraphics[scale=1.1,width=6.5cm]{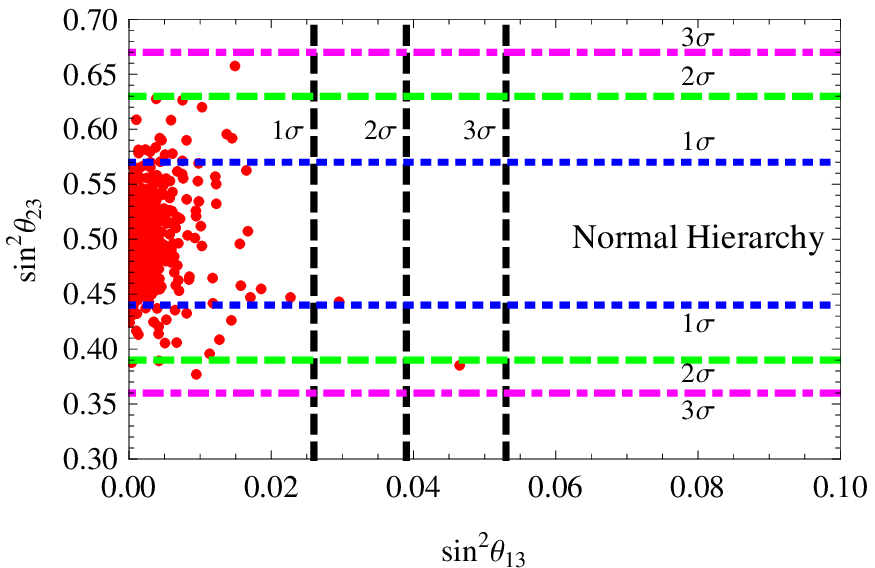}&\hspace{0.5cm}\includegraphics[scale=1.1,width=6.5cm]{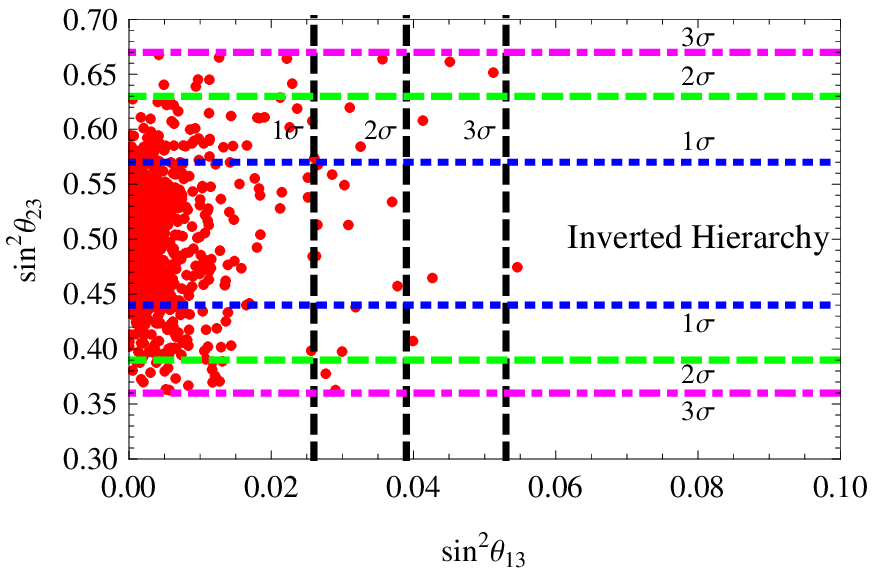}\\
~~~~~~(c)&~~~~~~~~~~(d)
\end{tabular}
\caption{\label{fig:mixing_angles} The allowed region of $\sin^2\theta_{12}$ and
$\sin^2\theta_{23}$ against $\sin^2\theta_{13}$ for both normal
hierarchy and inverted hierarchy neutrino mass spectrum}
\end{center}
\end{figure}

The analytic expressions for the mixing angles are shown in
Eq.(\ref{75}), which allows us to estimate the deviations from the
TB mixing pattern qualitatively. In order to see clearly the allowed
ranges of the mixing parameters, as well as for cross-checking the
reliability of the analytical estimates, we shall perform a
numerical analysis. All the involved LO and NLO coefficients are
taken to random complex number with absolute value in the interval
[$1/3$, 3], the expansion parameters $\varepsilon$ and $\epsilon$
have been fixed at the representative value of 0.04 and 0.001
respectively \footnote{For other indicative values of the small
parameters $\varepsilon$ and $\epsilon$, the resulting numerical
results don't change qualitatively.}, and the VEV ratio
$v_{\overline{45}}/v_{\overline{5}}$ is set to the indicative value
0.22. Furthermore, we require the oscillation parameters $\Delta
m^2_{21}$, $\Delta m^2_{31}$, $\sin^2\theta_{12}$,
$\sin^2\theta_{23}$ and $\sin^2\theta_{13}$ to lie in their
$3\sigma$ interval. The allowed regions of
$\sin^2\theta_{12}-\sin^2\theta_{13}$ and
$\sin^2\theta_{23}-\sin^2\theta_{13}$ for both normal hierarchy and
inverted hierarchy are showed in Fig. \ref{fig:mixing_angles}. It is
obvious that rather small $\theta_{13}$ is favored for both NH and
IH spectrum, which is consistent with our theoretical analysis.

\subsection{Neutrinoless double beta decay}
\begin{figure}[hptb]
\begin{center}
\includegraphics[scale=1.2]{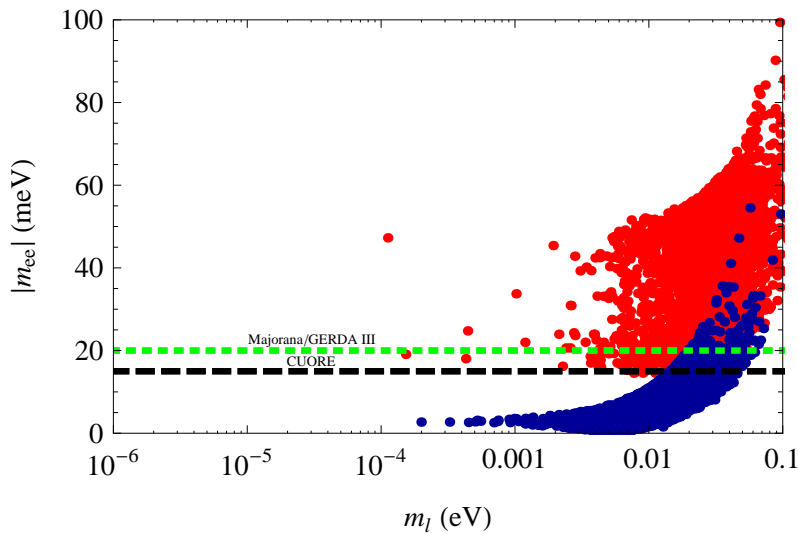}
\caption{\label{fig:m_ee} Scatter plot of the effective mass
$|m_{ee}|$ with respect to the lightest neutrino mass. The blue corresponds to
the normal hierarchy neutrino spectrum and the red to the inverted hierarchy spectrum.}
\end{center}
\end{figure}

The discovery of neutrinoless double beta decay $0\nu2\beta$ is very
important because it could directly establish lepton number
violation and the Majorana nature of neutrino. The $0\nu2\beta$
decay amplitude is proportional to the effective mass $|m_{ee}|$,
which is (11) element of the neutrino mass matrix in the basis where
the charged lepton mass matrix is real and diagonal. The allowed
region for $m_{ee}$ is displayed in Fig. \ref{fig:m_ee}. The
horizontal lines denote the future sensitivity of some $0\nu2\beta$
decay experiments, which are 15 meV and 20 meV respectively of CUORE
\cite{cuore}, Majorana \cite{majorana}/GERDA III \cite{gerda}
experiments.

We see that $|m_{ee}|$ is predicted be below the present
bound from the Heidelberg-Moscow experiment \cite{HM}. For the NH
spectrum, $|m_{ee}|$ can be so small to be close to zero, while the
scatter plot indicates a lower bound for $|m_{ee}|$ of about 14 meV
in the case of IH spectrum. It is quite close to the future experimental sensitivity so that $0\nu2\beta$ decay should be observed by future experiments for IH spectrum. We note that a partial cancellation between the LO and NLO
contributions takes place, so that the lightest neutrino mass can be
very small of order $10^{-4}$ eV in contrast with the LO constraints
shown in Eq.(\ref{36}). This however requires an additional fine
tuning of the parameters which has been reproduced in our numerical
analysis only partially and by very few points.

\begin{figure}[hptb]
\begin{center}
\includegraphics[scale=1.2]{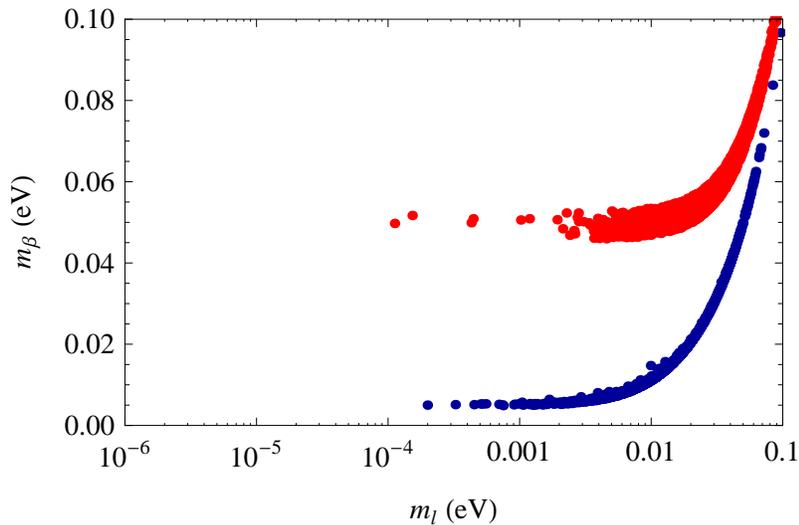}
\caption{\label{fig:mass_beta} $m_{\beta}$ as a function of the
lightest neutrino mass. The blue corresponds to
the normal hierarchy neutrino spectrum and the red to the inverted hierarchy spectrum.}
\end{center}
\end{figure}

In Fig.\ref{fig:mass_beta}, we plot the effective mass
$m_{\beta}=\big[\sum_k|(U_{PMNS})_{ek}|^2m^2_k\big]^{1/2}$ in
$\beta$ decay experiments, which could measure the non-zero neutrino
masses. We clearly see that the effective mass $m_{\beta}$ is
predicted to be below the prospective sensitivity 0.2 eV of the
KATRIN experiment for both NH and IH neutrino mass spectrum
\cite{katrin}.

\subsection{Sum of neutrino masses}
\begin{figure}[hptb]
\begin{center}
\includegraphics[scale=1.2]{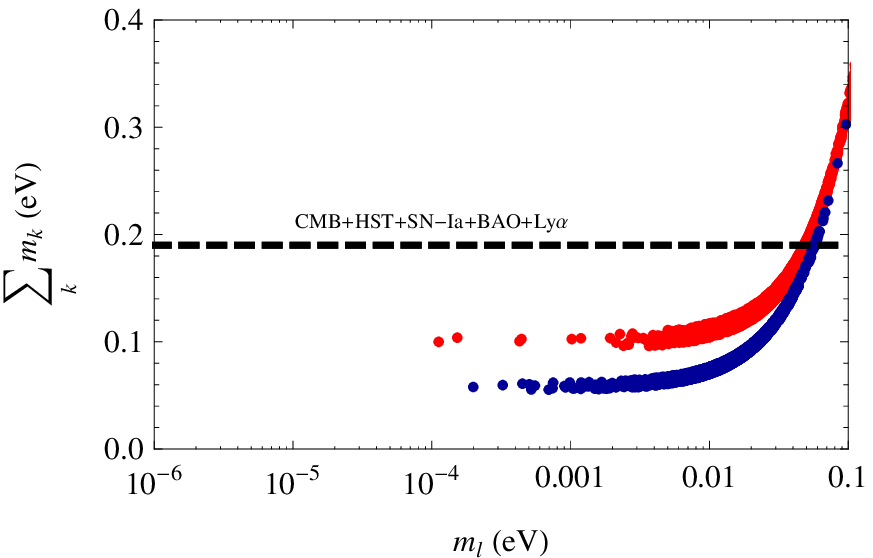}
\caption{\label{fig:mass_sum} The sum of neutrino masses $\sum_km_k$
as a function of the lightest neutrino mass. The blue corresponds to
the normal hierarchy neutrino spectrum and the red to the inverted hierarchy spectrum.  }
\end{center}
\end{figure}

The prediction for the sum of neutrino mass is presented in Fig.
\ref{fig:mass_sum}. The horizontal line is the cosmological bound at
$0.19$ eV, which is obtained by combining the data from the Cosmic
Microwave Background (CMB) anisotropy (from WMAP~5y \cite{WMAP2},
Arcminute Cosmology Bolometer Array Receiver (ACBAR) \cite{acbar07},
Very Small Array (VSA) \cite {vsa}, Cosmic Background Imager (CBI)
\cite{cbi} and BOOMERANG \cite{boom03} experiments) plus the
large-scale structure (LSS) information on galaxy clustering (from
the Luminous Red Galaxies Sloan Digital Sky Survey (SDSS)
\cite{Tegmark}) plus the Hubble Space Telescope (HST) plus the
luminosity distance SN-Ia data of \cite{astier} plus the BAO data
from \cite{bao} and finally plus the small scale primordial spectrum
from Lyman-alpha (Ly$\alpha$) forest clouds \cite{Ly1}. We see that
our model predicts $\sum_km_k$ too similar for both hierarchies to
be distinguished using the current cosmological information on the
sum of the neutrino masses.

The present model has rich phenomenological implications, we only
study few interesting observables here. Especially the predictions
for lepton flavor violation branching ratios and leptogenesis
deserve to be studied carefully, which are important to test the
model and distinguish this model from other discrete flavor symmetry
models. These topics will be discussed in future work \cite{ding}.

\section{Conclusion}
In this work, we have built a SUSY $SU(5)$ model based on the flavor
symmetry $S_4\times Z_3\times Z_4$. Three generations of adjoint
matter superfields are introduced, and they are assigned to
transform as $\mathbf{3_1}$ of $S_4$. The neutrino masses are
generated via the combination of type I and type III see-saw
mechanism in the model. To describe quarks, we make use of the
doublet representation of $S_4$, we accommodate the first two
generations of tenplets $\mathbf{10}$ in doublet under $S_4$. In
particular, the first generation $\mathbf{10}$ is assigned to the
second component of the doublet, and the second generation as the
first component, whereas the third generation of $\mathbf{10}$ is
kept invariant. The observed mass hierarchies of quarks are
reproduced naturally via the spontaneous breaking of the flavor
symmetry without invoking the Froggatt-Nielsen mechanism. In order
to generate the CKM mixing between the first two generations, we
require a moderate fine tuning
$v_{\overline{45}}/v_{\overline{5}}\sim\lambda_c$, then the model
generates the observed pattern of the CKM mixing matrix.

In the neutrino sector, the flavor symmetry $S_4$ is broken down to
the Klein four subgroup by the VEV of the flavon fields $\chi$ and
$\varphi$ at LO. The resulting light neutrino mass matrix is exactly
diagonalized by the TB mixing matrix, and the neutrino mass spectrum
can be normal hierarchy or inverted hierarchy. There are only three
independent parameters in the neutrino sector at LO, the model is
rather predictive so that the lightest neutrino mass is constrained
to be larger than 0.011 eV and 0.028 eV for normal hierarchy and
inverted hierarchy respectively. The mixing of the left-handed charged
leptons results in corrections to the TB mixing pattern, which is described in terms of a well-known
lepton mixing sum rule, and the reactor mixing angle is
predicted to be close to three degrees.

The subleading corrections to the flavon alignment and the fermion
mass matrices have been analyzed carefully. We show that the
successful LO predictions for the pattern of quark masses and CKM
mixing angles are not spoiled by the subleading contributions, and
all the three lepton mixing angles receive corrections of order
$\lambda^2_c$. The phenomenological implications of the model are
investigated in details, we find that future neutrinoless double
beta decay experiment with high precision is an important test of
the model, it allow us to distinguish the NH spectrum from the IH
one. Finally we note that the predictions presented in the work are
valid just below the GUT scale. To determine the fermion masses and
mixings and phenomenologically interesting observables at the
electroweak scale, we should study the renormalization group running
carefully. These issues will be studied elsewhere in future
\cite{ding}.


\section*{Acknowledgements}

We are grateful to Prof. Dao-Neng Gao and Mu-Lin Yan for stimulating
discussions. This work is supported by the National Natural Science
Foundation of China under Grant No.10905053, Chinese Academy
KJCX2-YW-N29 and the 973 project with Grant No. 2009CB825200.

\vfill
\newpage

\section*{Appendix A: The group $S_4$ and its representation}

The group $S_4$ is the group of the permutations of four objects, it
has $4!=24$ elements. Let a generic permutation be denoted by
$(1,2,3,4)\rightarrow(n_1,n_2,n_3,n_4)\equiv(n_1n_2n_3n_4)$. $S_4$
can be generated by the two basic permutations $S$ and $T$ where
$S=(2341)$ and $T=(2314)$. We can check that $S^{4}=T^3=1$,
$ST^2S=T$. $S_4$ has five conjugate classes as follows
\begin{eqnarray*}
&&{\cal C}_1:1\\
&&{\cal
C}_2:\;STS^2=(2134),\;TSTS^2=(3214),\;ST^2=(4231),\;S^2TS=(1324),\;TST=(1432),\\
&&~~~~~~T^2S=(1243)\\
&&{\cal C}_3:\;TS^2T^2=(2143),\;S^2=(3412),\;T^2S^2T=(4321)\\
&&{\cal
C}_4:\;T=(2314),\;T^2=(3124),\;T^2S^2=(2431),\;S^2T=(4132),\;S^2TS^2=(3241),\;STS=(4213),\\
&&~~~~~S^2T^2=(1342),\;TS^2=(1423)\\
&&{\cal
C}_5:\;S=(2341),\;T^2ST=(2413),\;ST=(3421),\;\;TS=(3142),\;TST^2=(4312),\;S^3=(4123)\\
\end{eqnarray*}
The structure of the group $S_4$ is rather rich, it has thirty
proper subgroups of orders 1, 2, 3, 4, 6, 8, 12 or 24. Concretely,
the details about the subgroups of $S_4$ can be found in Ref.
\cite{Ding:2009iy}. For a finite group the number of irreducible
representation is equal to the number of conjugate class.
Consequently the $S_4$ group have five irreducible representation:
$\mathbf{1_1}$, $\mathbf{1_2}$, $\mathbf{2}$, $\mathbf{3_1}$ and
$\mathbf{3_2}$, which are all real. Concretely the representation
matrix can be chosen as
\begin{eqnarray*}
\begin{array}{lcc}
&S=1, &T=1~~~~~\rm{for}~~~~~ \mathbf{1_1}\\
&S=-1,&T=1~~~~~\rm{for}~~~~~\mathbf{1_2}\\
&S=\left(\begin{array}{cc} 0&1\\
1&0
\end{array}
\right),&T=\left(\begin{array}{cc}
\omega&0\\
0&\omega^2
\end{array}
\right) ~~~\rm{for}~~~ \mathbf{2}
\\
&S=\frac{1}{3}\left(\begin{array}{ccc} -1&2\omega&2\omega^2\\
2\omega&2\omega^2&-1\\
2\omega^2&-1&2\omega
\end{array}
\right),&T=\left(\begin{array}{ccc}
1&0&0\\
0&\omega^2&0\\
0&0&\omega
\end{array}
\right)~~~\rm{for}~~~ \mathbf{3_1}\\
&S=\frac{-1}{3}\left(\begin{array}{ccc} -1&2\omega&2\omega^2\\
2\omega&2\omega^2&-1\\
2\omega^2&-1&2\omega
\end{array}
\right), &T=\left(\begin{array}{ccc}
1&0&0\\
0&\omega^2&0\\
0&0&\omega
\end{array}
\right)~~~\rm{for}~~~ \mathbf{3_2}
\end{array}
\end{eqnarray*}
The characters, i.e. the trace of the representation matrix, are
given in the character table (please see Table \ref{tab:character}).
From the character table of the $S_4$ group, we can
straightforwardly obtain the multiplication rules between the
various representations
\begin{eqnarray}
\nonumber&&\mathbf{1}_i\otimes\mathbf{1}_j=\mathbf{1}_{((i+j)\;{\rm
mod}\;
2)+1},~~~~\mathbf{1}_i\otimes\mathbf{2}=\mathbf{2},~~~~\mathbf{1}_i\otimes\mathbf{3}_j=\mathbf{3}_{((i+j)\;{\rm
mod}\;
2)+1}\\
\nonumber&&\mathbf{2}\otimes\mathbf{2}=\mathbf{1_1}\oplus\mathbf{1_2}\oplus\mathbf{2},~~~~\mathbf{2}\otimes\mathbf{3}_i=\mathbf{3_1}\oplus\mathbf{3_2},~~~~\mathbf{3}_i\otimes\mathbf{3}_i=\mathbf{1_1}\oplus\mathbf{2}\oplus\mathbf{3_1}\oplus\mathbf{3_2},\\
\label{ap1}&&\mathbf{3_1}\otimes\mathbf{3_2}=\mathbf{1_2}\oplus\mathbf{2}\oplus\mathbf{3_1}\oplus\mathbf{3_2},
~~~{\rm with}~ i,j=1,2
\end{eqnarray}
\begin{table}[ht]
\begin{center}
\begin{tabular}{|c|c|c|c|c|c|}\hline\hline
   &\multicolumn{5}{|c|}{Classes}\\\cline{2-6}
   &${\cal C}_1$&${\cal C}_2$&${\cal C}_3$&${\cal C}_4$&${\cal
C}_5$\\\hline



$\mathbf{1_1}$&1&1&1&1&1\\\hline

$\mathbf{1_2}$&1&-1&1&1&-1\\\hline

$\mathbf{2}$&2&0&2&-1&0\\\hline

$\mathbf{3_1}$&3&1&-1&0&-1\\\hline

$\mathbf{3_2}$&3&-1&-1&0&1\\\hline\hline
\end{tabular}
\caption{\label{tab:character}Character table of the $S_4$ group.}
\end{center}
\end{table}
Starting from the explicit matrix representation, we get the product
decomposition rules of the $S_4$ group. In the following we use
$\alpha_i$ to denote the elements of the first representation of the
product and $\beta_i$ to indicate those of the second
representation.
\begin{itemize}
\item{$\mathbf{1_2}\otimes\mathbf{1_2}=\mathbf{1_1}$}
\begin{equation}
\label{ap2}\mathbf{1_1}\sim \alpha\beta
\end{equation}
\item{$\mathbf{1_2}\otimes\mathbf{2}=\mathbf{2}$}
\begin{equation}
\label{ap3}\mathbf{2}\sim\left(\begin{array}{c}
\alpha\beta_1\\
-\alpha\beta_2\end{array}\right)
\end{equation}
\item{$\mathbf{1_2}\otimes\mathbf{3_1}=\mathbf{3_2}$}
\begin{equation}
\label{ap4}\mathbf{3_2}\sim\left(\begin{array}{c}
\alpha\beta_1\\
\alpha\beta_2\\
\alpha\beta_3\end{array}\right)
\end{equation}
\item{$\mathbf{1_2}\otimes\mathbf{3_2}=\mathbf{3_1}$}
\begin{equation}
\label{ap5}\mathbf{3_1}\sim\left(\begin{array}{c}
\alpha\beta_1\\
\alpha\beta_2\\
\alpha\beta_3\end{array}\right)
\end{equation}
\item{$\mathbf{2}\otimes\mathbf{2}=\mathbf{1_1}\oplus\mathbf{1_2}\oplus\mathbf{2}$}
\begin{equation}
\label{ap6}\mathbf{1_1}\sim \alpha_1\beta_2+\alpha_2\beta_1
\end{equation}
\begin{equation}
\label{ap7}\mathbf{1_2}\sim \alpha_1\beta_2-\alpha_2\beta_1
\end{equation}
\begin{equation}
\label{ap8}\mathbf{2}\sim
\left(\begin{array}{c}\alpha_2\beta_2\\
\alpha_1\beta_1\end{array}\right)
\end{equation}
\item{$\mathbf{3_1}\otimes\mathbf{3_1}=\mathbf{3_2}\otimes\mathbf{3_2}=\mathbf{1_1}\oplus\mathbf{2}\oplus\mathbf{3_1}\oplus\mathbf{3_2}$}
\begin{equation}
\label{ap9}\mathbf{1_1}\sim\alpha_1\beta_1+\alpha_2\beta_3+\alpha_3\beta_2
\end{equation}
\begin{equation}
\label{ap10}\mathbf{2}\sim\left(\begin{array}{c}
\alpha_2\beta_2+\alpha_1\beta_3+\alpha_3\beta_1\\
\alpha_3\beta_3+\alpha_1\beta_2+\alpha_2\beta_1
\end{array}\right)
\end{equation}
\begin{equation}
\label{ap11}\mathbf{3_1}\sim\left(\begin{array}{c}
2\alpha_1\beta_1-\alpha_2\beta_3-\alpha_3\beta_2\\
2\alpha_3\beta_3-\alpha_1\beta_2-\alpha_2\beta_1\\
2\alpha_2\beta_2-\alpha_1\beta_3-\alpha_3\beta_1
\end{array}\right)
\end{equation}
\begin{equation}
\label{ap12}\mathbf{3_2}\sim\left(\begin{array}{c}
\alpha_2\beta_3-\alpha_3\beta_2\\
\alpha_1\beta_2-\alpha_2\beta_1\\
\alpha_3\beta_1-\alpha_1\beta_3
\end{array}\right)
\end{equation}
\item{$\mathbf{3_1}\otimes\mathbf{3_2}=\mathbf{1_2}\oplus\mathbf{2}\oplus\mathbf{3_1}\oplus\mathbf{3_2}$}
\begin{equation}
\label{ap13}\mathbf{1_2}\sim
\alpha_1\beta_1+\alpha_2\beta_3+\alpha_3\beta_2
\end{equation}
\begin{equation}
\label{ap14}\mathbf{2}\sim\left(\begin{array}{c}\alpha_2\beta_2+\alpha_1\beta_3+\alpha_3\beta_1\\
-\alpha_3\beta_3-\alpha_1\beta_2-\alpha_2\beta_1\end{array}\right)
\end{equation}
\begin{equation}
\label{ap15}\mathbf{3_1}\sim\left(\begin{array}{c}
\alpha_2\beta_3-\alpha_3\beta_2\\
\alpha_1\beta_2-\alpha_2\beta_1\\
\alpha_3\beta_1-\alpha_1\beta_3
\end{array}\right)
\end{equation}
\begin{equation}
\label{ap16}\mathbf{3_2}\sim\left(\begin{array}{c}
2\alpha_1\beta_1-\alpha_2\beta_3-\alpha_3\beta_2\\
2\alpha_3\beta_3-\alpha_1\beta_2-\alpha_2\beta_1\\
2\alpha_2\beta_2-\alpha_1\beta_3-\alpha_3\beta_1
\end{array}\right)
\end{equation}
\item{$\mathbf{2}\otimes\mathbf{3_1}=\mathbf{3_1}\oplus\mathbf{3_2}$}
\begin{equation}
\label{ap17}\mathbf{3_1}\sim\left(\begin{array}{c}
\alpha_1\beta_2+\alpha_2\beta_3\\
\alpha_1\beta_3+\alpha_2\beta_1\\
\alpha_1\beta_1+\alpha_2\beta_2
\end{array}\right)
\end{equation}
\begin{equation}
\label{ap18}\mathbf{3_2}\sim\left(\begin{array}{c}
\alpha_1\beta_2-\alpha_2\beta_3\\
\alpha_1\beta_3-\alpha_2\beta_1\\
\alpha_1\beta_1-\alpha_2\beta_2
\end{array}\right)
\end{equation}
\item{$\mathbf{2}\otimes\mathbf{3_2}=\mathbf{3_1}\oplus\mathbf{3_2}$}
\begin{equation}
\label{ap19}\mathbf{3_1}\sim\left(\begin{array}{c}
\alpha_1\beta_2-\alpha_2\beta_3\\
\alpha_1\beta_3-\alpha_2\beta_1\\
\alpha_1\beta_1-\alpha_2\beta_2
\end{array}\right)
\end{equation}
\begin{equation}
\label{ap20}\mathbf{3_2}\sim\left(\begin{array}{c}
\alpha_1\beta_2+\alpha_2\beta_3\\
\alpha_1\beta_3+\alpha_2\beta_1\\
\alpha_1\beta_1+\alpha_2\beta_2
\end{array}\right)
\end{equation}
\end{itemize}
We note that the multiplication rules
presented above are in accordance with the results in Ref.
\cite{Bazzocchi:2009pv}.

\section*{Appendix B: Vacuum alignment beyond the leading order}
In this appendix we discuss the subleading terms of the driving
superpotential $w_v$ and the resulting corrections to the LO vacuum
alignment. At the next level of approximation the LO driving
superpotential is corrected by operators of higher dimension whose
contributions are suppressed by at least one power of $1/\Lambda$.
As a result, the superpotential depending on the driving fields
becomes
\begin{equation}
\label{ap21}w_v=w^0_v+\delta w_v
\end{equation}
The leading order term $w^0_v$ reads
\begin{eqnarray}
\nonumber
w^0_v&=&f_1\chi^0(\chi\varphi)_{3_2}+f_2\chi^0\chi\zeta+f_3\varphi^0(\chi\chi)_2+f_4\varphi^0(\varphi\varphi)_2+f_5\varphi^0\varphi\zeta
+g_1\phi^0(\phi\phi)_{3_1}\\
\label{app22}&&+g_2\phi^0(\eta\phi)_{3_1}+g_3\rho^0(\phi\phi)_{1_1}+g_4\rho^0(\eta\eta)_{1_1}+h_1\Delta^0(\Delta\Delta)_{3_1}+h_2\Delta^0\Delta\xi
\end{eqnarray}
In the SUSY limit, $w^0_v$ leads to the following vacuum
configuration
\begin{eqnarray}
\nonumber&&\langle\chi\rangle=\left(\begin{array}{c}v_{\chi}\\v_{\chi}\\v_{\chi}\end{array}\right),~~~\langle\varphi\rangle=\left(\begin{array}{c}v_{\varphi}\\
v_{\varphi}\end{array}\right),~~~\langle\zeta\rangle=0,~~~{\rm
with}~v^2_{\chi}=-\frac{f_4}{3f_3}v^2_{\varphi}\\
\nonumber&&\langle\phi\rangle=\left(\begin{array}{c}0\\
v_{\phi}\\0\end{array}\right),~~~\langle\eta\rangle=\left(\begin{array}{c}0\\v_{\eta}\end{array}\right),~~~{\rm
with}~v_{\phi}=-\frac{g_2}{2g_1}v_{\eta}\\
\label{ap23}&&\langle\Delta\rangle=\left(\begin{array}{c}v_{\Delta}\\0\\0\end{array}\right),~~~\langle\xi\rangle=v_{\xi},~~~{\rm
with}~v_{\Delta}=-\frac{h_2}{2h_1}v_{\xi}
\end{eqnarray}
The correction term $\delta w_v$ consists of the most general
subleading operators linear in the driving fields, and they should
be invariant under the flavor symmetry $S_4\times Z_3\times Z_4$.
\begin{equation}
\label{ap24}\delta w_v=\frac{1}{\Lambda}\sum^{15}_{i=1}k_i{\cal
O}^{\chi^0}_i+\frac{1}{\Lambda}\sum^{10}_{i=1}w_i{\cal
O}^{\varphi^0}_i+\frac{1}{\Lambda^2}\sum^{39}_{i=1}s_i{\cal
O}^{\phi^0}_i+\frac{1}{\Lambda^2}\sum^{18}_{i=1}r_i{\cal
O}^{\rho^0}_i+\frac{1}{\Lambda^2}\sum^{28}_{i=1}t_i{\cal
O}^{\Delta^0}_i
\end{equation}
where $k_i$, $w_i$, $s_i$, $r_i$ and
$t_i$ are order one coefficients,
their specific values are not determined by the flavor symmetry.
\{${\cal O}^{\chi^0}_i$, ${\cal O}^{\varphi^0}_i$, ${\cal
O}^{\phi^0}_i$, ${\cal O}^{\rho^0}_i$, ${\cal O}^{\Delta^0}_i$\} denote the complete set of
subleading contractions invariant under $S_4\times Z_3\times Z_4$.
\begin{eqnarray}
\nonumber&&{\cal O}^{\chi^0}_1=(\chi^0\chi)_2(\phi\Delta)_2,~~{\cal
O}^{\chi^0}_2=(\chi^0\chi)_{3_1}(\phi\Delta)_{3_1},~~{\cal
O}^{\chi^0}_3=(\chi^0\chi)_{3_2}(\phi\Delta)_{3_2},\\
\nonumber&&{\cal
O}^{\chi^0}_4=(\chi^0\varphi)_{3_1}(\phi\Delta)_{3_1},~~{\cal
O}^{\chi^0}_5=(\chi^0\varphi)_{3_2}(\phi\Delta)_{3_2},~~{\cal
O}^{\chi^0}_6=(\chi^0\chi)_{3_1}\phi\xi,\\
\nonumber&&{\cal O}^{\chi^0}_7=(\chi^0\varphi)_{3_1}\phi\xi,~~{\cal
O}^{\chi^0}_8=(\chi^0\chi)_{3_1}(\eta\Delta)_{3_1},~~{\cal
O}^{\chi^0}_9=(\chi^0\chi)_{3_2}(\eta\Delta)_{3_2},\\
\nonumber&&{\cal
O}^{\chi^0}_{10}=(\chi^0\varphi)_{3_1}(\eta\Delta)_{3_1},~~{\cal
O}^{\chi^0}_{11}=(\chi^0\varphi)_{3_2}(\eta\Delta)_{3_2},~~{\cal
O}^{\chi^0}_{12}=(\chi^0\chi)_2\eta\xi\\
\label{ap25}&&{\cal
O}^{\chi^0}_{13}=\chi^{0}\zeta(\phi\Delta)_{3_1},~~{\cal
O}^{\chi^0}_{14}=\chi^0\zeta\phi\xi,~~{\cal
O}^{\chi^0}_{15}=\chi^0\zeta(\eta\Delta)_{3_1}
\end{eqnarray}
\begin{eqnarray}
\nonumber&&{\cal
O}^{\varphi^0}_1=(\varphi^0\chi)_{3_1}(\phi\Delta)_{3_1},~~{\cal
O}^{\varphi^0}_2=(\varphi^0\chi)_{3_2}(\phi\Delta)_{3_2},~~{\cal
O}^{\varphi^0}_3=(\varphi^0\varphi)_{1_1}(\phi\Delta)_{1_1},\\
\nonumber&&{\cal
O}^{\varphi^0}_4=(\varphi^0\varphi)_2(\phi\Delta)_2,~~ {\cal
O}^{\varphi^0}_5=(\varphi^0\chi)_{3_1}\phi\xi,~~{\cal
O}^{\varphi^0}_6=(\varphi^0\chi)_{3_1}(\eta\Delta)_{3_1},\\
\nonumber&&{\cal
O}^{\varphi^0}_7=(\varphi^0\chi)_{3_2}(\eta\Delta)_{3_2},~~{\cal
O}^{\varphi^0}_8=(\varphi^0\varphi)_2\eta\xi,~~{\cal
O}^{\varphi^0}_9=\varphi^0\zeta(\phi\Delta)_2\\
\label{ap26}&&{\cal O}^{\varphi^0}_{10}=\varphi^0\zeta\eta\xi
\end{eqnarray}
\begin{eqnarray}
\nonumber&&{\cal
O}^{\phi^0}_1=\phi^0(\chi\chi)_{1_1}(\phi\phi)_{3_1},~~{\cal
O}^{\phi^0}_2=\phi^0((\chi\chi)_2(\phi\phi)_{3_1})_{3_1},~~{\cal
O}^{\phi^0}_3=\phi^0(\chi\chi)_{3_1}(\phi\phi)_{1_1},\\
\nonumber&&{\cal
O}^{\phi^0}_4=\phi^0((\chi\chi)_{3_1}(\phi\phi)_2)_{3_1},~~{\cal
O}^{\phi^0}_5=\phi^0((\chi\chi)_{3_1}(\phi\phi)_{3_1})_{3_1},~~{\cal
O}^{\phi^0}_6=\phi^0(\chi\chi)_{1_1}(\eta\phi)_{3_1},\\
\nonumber&&{\cal
O}^{\phi^0}_7=\phi^0((\chi\chi)_2(\eta\phi)_{3_1})_{3_1},~~{\cal
O}^{\phi^0}_8=\phi^0((\chi\chi)_2(\eta\phi)_{3_2})_{3_1},~~{\cal
O}^{\phi^0}_9=\phi^0((\chi\chi)_{3_1}(\eta\phi)_{3_1})_{3_1},\\
\nonumber&&{\cal
O}^{\phi^0}_{10}=\phi^0((\chi\chi)_{3_1}(\eta\phi)_{3_2})_{3_1},~~{\cal
O}^{\phi^0}_{11}=\phi^0(\chi\chi)_{3_1}(\eta\eta)_{1_1},~~{\cal
O}^{\phi^0}_{12}=\phi^0((\chi\chi)_{3_1}(\eta\eta)_{2})_{3_1},\\
\nonumber&&{\cal
O}^{\phi^0}_{13}=\phi^0(\varphi\varphi)_{1_1}(\phi\phi)_{3_1},~~{\cal
O}^{\phi^0}_{14}=\phi^0((\varphi\varphi)_2(\phi\phi)_{3_1})_{3_1},~~{\cal
O}^{\phi^0}_{15}=\phi^0(\varphi\varphi)_{1_1}(\eta\phi)_{3_1},\\
\nonumber&&{\cal
O}^{\phi^0}_{16}=\phi^0((\varphi\varphi)_2(\eta\phi)_{3_1})_{3_1},~~{\cal
O}^{\phi^0}_{17}=\phi^0((\varphi\varphi)_2(\eta\phi)_{3_2})_{3_1},~~{\cal
O}^{\phi^0}_{18}=\phi^0(\varphi\chi)_{3_1}(\phi\phi)_{1_1},\\
\nonumber&&{\cal
O}^{\phi^0}_{19}=\phi^0((\varphi\chi)_{3_1}(\phi\phi)_2)_{3_1},~~{\cal
O}^{\phi^0}_{20}=\phi^0((\varphi\chi)_{3_1}(\phi\phi)_{3_1})_{3_1},~~{\cal
O}^{\phi^0}_{21}=\phi^0((\varphi\chi)_{3_2}(\phi\phi)_2)_{3_1},\\
\nonumber&&{\cal
O}^{\phi^0}_{22}=\phi^0((\varphi\chi)_{3_2}(\phi\phi)_{3_1})_{3_1},~~{\cal
O}^{\phi^0}_{23}=\phi^0((\varphi\chi)_{3_1}(\eta\phi)_{3_1})_{3_1},~~{\cal
O}^{\phi^0}_{24}=\phi^0((\varphi\chi)_{3_1}(\eta\phi)_{3_2})_{3_1},\\
\nonumber&&{\cal
O}^{\phi^0}_{25}=\phi^0((\varphi\chi)_{3_2}(\eta\phi)_{3_1})_{3_1},~~{\cal
O}^{\phi^0}_{26}=\phi^0((\varphi\chi)_{3_2}(\eta\phi)_{3_2})_{3_1},~~{\cal
O}^{\phi^0}_{27}=\phi^0(\varphi\chi)_{3_1}(\eta\eta)_{1_1},\\
\nonumber&&{\cal
O}^{\phi^0}_{28}=\phi^0((\varphi\chi)_{3_1}(\eta\eta)_2)_{3_1},~~{\cal
O}^{\phi^0}_{29}=\phi^0((\varphi\chi)_{3_2}(\eta\eta)_2)_{3_1},~~{\cal
O}^{\phi^0}_{30}=\phi^0(\chi(\phi\phi)_2)_{3_2}\zeta\\
\nonumber&&{\cal
O}^{\phi^0}_{31}=\phi^0(\chi(\phi\phi)_{3_1})_{3_2}\zeta,~~{\cal
O}^{\phi^0}_{32}=\phi^0(\chi(\eta\phi)_{3_1})_{3_2}\zeta,~~{\cal
O}^{\phi^0}_{33}=\phi^0(\chi(\eta\phi)_{3_2})_{3_2}\zeta,\\
\nonumber&&{\cal
O}^{\phi^0}_{34}=\phi^0(\chi(\eta\eta)_2)_{3_2}\zeta,~~{\cal
O}^{\phi^0}_{35}=\phi^0(\varphi(\phi\phi)_{3_1})_{3_2}\zeta,~~{\cal
O}^{\phi^0}_{36}=\phi^0(\varphi(\eta\phi)_{3_1})_{3_2}\zeta\\
\label{ap27}&&{\cal
O}^{\phi^0}_{37}=\phi^0(\varphi(\eta\phi)_{3_2})_{3_2}\zeta,~~{\cal
O}^{\phi^0}_{38}=\phi^0(\phi\phi)_{3_1}\zeta^2,~~{\cal
O}^{\phi^0}_{39}=\phi^0(\eta\phi)_{3_1}\zeta^2
\end{eqnarray}
\begin{eqnarray}
\nonumber&&{\cal O}^{\rho^0}_1=\rho^0(\chi\chi)_{1_1}(\phi\phi)_{1_1},~~{\cal O}^{\rho^0}_2=\rho^0(\chi\chi)_2(\phi\phi)_2,~~{\cal O}^{\rho^0}_3=\rho^0(\chi\chi)_{3_1}(\phi\phi)_{3_1},\\
\nonumber&&{\cal O}^{\rho^0}_4=\rho^0(\chi\chi)_{3_1}(\eta\phi)_{3_1},~~{\cal O}^{\rho^0}_5=\rho^0(\chi\chi)_{1_1}(\eta\eta)_{1_1},~~{\cal O}^{\rho^0}_6=\rho^0(\chi\chi)_2(\eta\eta)_2,\\
\nonumber&&{\cal O}^{\rho^0}_7=\rho^0(\varphi\varphi)_{1_1}(\phi\phi)_{1_1},~~{\cal O}^{\rho^0}_8=\rho^0(\varphi\varphi)_2(\phi\phi)_2,~~{\cal O}^{\rho^0}_9=\rho^0(\varphi\varphi)_{1_1}(\eta\eta)_{1_1},\\
\nonumber&&{\cal O}^{\rho^0}_{10}=\rho^0(\varphi\varphi)_2(\eta\eta)_2,~~{\cal O}^{\rho^0}_{11}=\rho^0(\varphi\chi)_{3_1}(\phi\phi)_{3_1},~~{\cal O}^{\rho^0}_{12}=\rho^0(\varphi\chi)_{3_1}(\eta\phi)_{3_1},\\
\nonumber&&{\cal O}^{\rho^0}_{13}=\rho^0(\varphi\chi)_{3_2}(\eta\phi)_{3_2},~~{\cal O}^{\rho^0}_{14}=\rho^0\chi\zeta(\eta\phi)_{3_2},~~{\cal O}^{\rho^0}_{15}=\rho^0\varphi\zeta(\phi\phi)_2,\\
\label{add1}&&{\cal O}^{\rho^0}_{16}=\rho^0\varphi\zeta(\eta\eta)_2,~~{\cal O}^{\rho^0}_{17}=\rho^0\zeta^2(\phi\phi)_{1_1},~~{\cal O}^{\rho^0}_{18}=\rho^0\zeta^2(\eta\eta)_{1_1}
\end{eqnarray}
\begin{eqnarray}
\nonumber&&{\cal
O}^{\Delta^0}_1=\Delta^0(\chi\chi)_{1_1}(\Delta\Delta)_{3_1},~~{\cal
O}^{\Delta^0}_2=\Delta^0((\chi\chi)_2(\Delta\Delta)_{3_1})_{3_1},~~{\cal
O}^{\Delta^0}_3=\Delta^0(\chi\chi)_{3_1}(\Delta\Delta)_{1_1},\\
\nonumber&&{\cal
O}^{\Delta^0}_4=\Delta^0((\chi\chi)_{3_1}(\Delta\Delta)_2)_{3_1},~~{\cal
O}^{\Delta^0}_5=\Delta^0((\chi\chi)_{3_1}(\Delta\Delta)_{3_1})_{3_1},~~{\cal
O}^{\Delta^0}_6=\Delta^0(\chi\chi)_{1_1}\Delta\xi,\\
\nonumber&&{\cal
O}^{\Delta^0}_7=\Delta^0((\chi\chi)_2\Delta)_{3_1}\xi,~~{\cal
O}^{\Delta^0}_8=\Delta^0((\chi\chi)_{3_1}\Delta)_{3_1}\xi,~~{\cal
O}^{\Delta^0}_9=\Delta^0(\chi\chi)_{3_1}\xi\xi,\\
\nonumber&&{\cal
O}^{\Delta^0}_{10}=\Delta^0(\varphi\chi)_{3_1}(\Delta\Delta)_{1_1},~~{\cal
O}^{\Delta^0}_{11}=\Delta^0((\varphi\chi)_{3_1}(\Delta\Delta)_2)_{3_1},~~{\cal
O}^{\Delta^0}_{12}=\Delta^0((\varphi\chi)_{3_1}(\Delta\Delta)_{3_1})_{3_1},\\
\nonumber&&{\cal
O}^{\Delta^0}_{13}=\Delta^0((\varphi\chi)_{3_2}(\Delta\Delta)_2)_{3_1},~~{\cal
O}^{\Delta^0}_{14}=\Delta^0((\varphi\chi)_{3_2}(\Delta\Delta)_{3_1})_{3_1},~~{\cal
O}^{\Delta^0}_{15}=\Delta^0((\varphi\chi)_{3_1}\Delta)_{3_1}\xi,\\
\nonumber&&{\cal
O}^{\Delta^0}_{16}=\Delta^0((\varphi\chi)_{3_2}\Delta)_{3_1}\xi,~~{\cal
O}^{\Delta^0}_{17}=\Delta^0(\varphi\chi)_{3_1}\xi\xi,~~{\cal
O}^{\Delta^0}_{18}=\Delta^0(\varphi\varphi)_{1_1}(\Delta\Delta)_{3_1},\\
\nonumber&&{\cal
O}^{\Delta^0}_{19}=\Delta^0((\varphi\varphi)_2(\Delta\Delta)_{3_1})_{3_1},~~{\cal
O}^{\Delta^0}_{20}=\Delta^0(\varphi\varphi)_{1_1}\Delta\xi,~~{\cal
O}^{\Delta^0}_{21}=\Delta^0((\varphi\varphi)_2\Delta)_{3_1}\xi,\\
\nonumber&&{\cal
O}^{\Delta^0}_{22}=\Delta^0(\chi(\Delta\Delta)_2)_{3_2}\zeta,~~{\cal
O}^{\Delta^0}_{23}=\Delta^0(\chi(\Delta\Delta)_{3_1})_{3_2}\zeta,~~{\cal
O}^{\Delta^0}_{24}=\Delta^0(\chi\Delta)_{3_2}\xi\zeta,\\
\nonumber&&{\cal
O}^{\Delta^0}_{25}=\Delta^0(\varphi(\Delta\Delta)_{3_1})_{3_2}\zeta,~~{\cal
O}^{\Delta^0}_{26}=\Delta^0(\varphi\Delta)_{3_2}\xi\zeta,~~{\cal
O}^{\Delta^0}_{27}=\Delta^0(\Delta\Delta)_{3_1}\zeta^2,\\
\label{ap28}&&{\cal
O}^{\Delta^0}_{28}=\Delta^0\Delta\xi\zeta^2
\end{eqnarray}
The subleading contribution $\delta w_v$ induces shifts in the LO
VEVs shown above, then the new vacuum configuration can be
parameterized as
\begin{eqnarray}
\nonumber&&\langle\chi\rangle=\left(\begin{array}{c}v_{\chi}+\delta v_{\chi1}\\v_{\chi}+\delta v_{\chi2}\\v_{\chi}+\delta v_{\chi3}\end{array}\right),~~~~~\langle\varphi\rangle=\left(\begin{array}{c}v_{\varphi}\\
v_{\varphi}+\delta v_{\varphi2}\end{array}\right),~~~~~\langle\zeta\rangle=\delta v_{\zeta}\\
\nonumber&&\langle\phi\rangle=\left(\begin{array}{c}\delta v_{\phi1}\\
v_{\phi}+\delta v_{\phi2}\\ \delta v_{\phi3}\end{array}\right),~~~~~\langle\eta\rangle=\left(\begin{array}{c}\delta
v_{\eta1} \\v_{\eta}\end{array}\right)\\
\label{ap29}&&\langle\Delta\rangle=\left(\begin{array}{c}v_{\Delta}+\delta
v_{\Delta1}\\ \delta v_{\Delta2}\\ \delta
v_{\Delta3}\end{array}\right),~~~~~\langle\xi\rangle=v_{\xi}
\end{eqnarray}
where the shifts $\delta v_{\varphi1}$, $\delta v_{\eta2}$ and
$\delta v_{\xi}$ have been absorbed into the redefinition of the undetermined parameters
$v_{\varphi}$, $v_{\eta}$ and $v_{\xi}$ respectively. The new vacua is obtained by searching for the zeros of the F-terms,
i.e. the first derivative of $w_v+\delta w_v$ with respect to the
driving fields $\chi^0$, $\varphi^0$, $\phi^0$, $\rho^0$ and $\Delta^0$. By
keeping only the terms linear in the shift $\delta v$ and neglecting
the terms proportional to $\delta v/\Lambda$, the minimization
equations become
\begin{eqnarray}
\nonumber&&f_1[v_{\varphi}(\delta v_{\chi2}-\delta
v_{\chi3})-v_{\chi}\delta
v_{\varphi2}]+f_2v_{\chi}\delta v_{\zeta}+a_1v_{\chi}v_{\phi}v_{\Delta}/\Lambda=0\\
\nonumber&&f_1[v_{\varphi}(\delta v_{\chi1}-\delta
v_{\chi2})-v_{\chi}\delta
v_{\varphi2}]+f_2v_{\chi}\delta v_{\zeta}+a_2v_{\chi}v_{\phi}v_{\Delta}/\Lambda=0\\
\nonumber&&f_1[v_{\varphi}(\delta v_{\chi3}-\delta
v_{\chi1})-v_{\chi}\delta
v_{\varphi2}]+f_2v_{\chi}\delta v_{\zeta}+a_3v_{\chi}v_{\phi}v_{\Delta}/\Lambda=0\\
\nonumber&&2f_3v_{\chi}(\delta v_{\chi1}+\delta v_{\chi2}+\delta
v_{\chi3})+f_5v_{\varphi}\delta v_{\zeta}+b_1v_{\chi}v_{\phi}v_{\Delta}/\Lambda=0\\
\label{ap30}&&2f_3v_{\chi}(\delta v_{\chi1}+\delta v_{\chi2}+\delta
v_{\chi3})+2f_4v_{\varphi}\delta
v_{\varphi2}-f_5v_{\varphi}\delta v_{\zeta}+b_2v_{\chi}v_{\phi}v_{\Delta}/\Lambda=0
\end{eqnarray}
where the coefficients $a_{1,2,3}$ and $b_{1,2}$ are linear
combinations of the subleading coefficients
\begin{eqnarray}
\nonumber a_1&=&k_1+k_2+k_3-(k_4+k_5)v_{\varphi}/v_{\chi}-k_6v_{\xi}/v_{\Delta}+k_7v_{\varphi}v_{\xi}/(v_{\chi}v_{\Delta})+(-k_8+k_9)v_{\eta}/v_{\phi}\\
\nonumber&&+(k_{10}-k_{11})v_{\varphi}v_{\eta}/(v_{\chi}v_{\phi})+k_{12}v_{\eta}v_{\xi}/(v_{\phi}v_{\Delta})\\
\nonumber
a_2&=&k_1-2k_3+(k_4-k_5)v_{\varphi}/v_{\chi}-k_7v_{\varphi}v_{\xi}/(v_{\chi}v_{\Delta})-2k_9v_{\eta}/v_{\phi}-(k_{10}+k_{11})v_{\varphi}v_{\eta}/(v_{\chi}v_{\phi})\\
\nonumber&&+k_{12}v_{\eta}v_{\xi}/(v_{\phi}v_{\Delta})\\
\nonumber
a_3&=&k_1-k_2+k_3+k_6v_{\xi}/v_{\Delta}+(k_8+k_9)v_{\eta}/v_{\phi}+k_{12}v_{\eta}v_{\xi}/(v_{\phi}v_{\Delta})\\
\nonumber
b_1&=&-w_1-w_2+w_5v_{\xi}/v_{\Delta}+(w_6-w_7)v_{\eta}/v_{\phi}\\
\label{ap31}
b_2&=&-w_1+w_2+w_4v_{\varphi}/v_{\chi}+w_5v_{\xi}/v_{\Delta}+(w_6+w_7)v_{\eta}/v_{\phi}+w_8v_{\varphi}v_{\eta}v_{\xi}/(v_{\chi}v_{\phi}v_{\Delta})
\end{eqnarray}
The solution to the linear equations Eq.(\ref{ap30}) is
\begin{eqnarray}
\nonumber&&\delta
v_{\chi1}=\frac{a_3-a_2}{3f_1}\frac{v_{\chi}v_{\phi}v_{\Delta}}{\Lambda
v_{\varphi}}-\frac{(a_1+a_2+a_3)f_4f_5}{18(f_1f_5-f_2f_4)f_3}\frac{v_{\varphi}v_{\phi}v_{\Delta}}{\Lambda v_{\chi}}-\frac{(b_1+b_2)f_1f_5-2b_1f_2f_4}{12(f_1f_5-f_2f_4)f_3}\frac{v_{\phi}v_{\Delta}}{\Lambda}\\
\nonumber&&\delta
v_{\chi2}=\frac{a_2-a_1}{3f_1}\frac{v_{\chi}v_{\phi}v_{\Delta}}{\Lambda
v_{\varphi}}-\frac{(a_1+a_2+a_3)f_4f_5}{18(f_1f_5-f_2f_4)f_3}\frac{v_{\varphi}v_{\phi}v_{\Delta}}{\Lambda v_{\chi}}-\frac{(b_1+b_2)f_1f_5-2b_1f_2f_4}{12(f_1f_5-f_2f_4)f_3}\frac{v_{\phi}v_{\Delta}}{\Lambda}\\
\nonumber&&\delta
v_{\chi3}=\frac{a_1-a_3}{3f_1}\frac{v_{\chi}v_{\phi}v_{\Delta}}{\Lambda
v_{\varphi}}-\frac{(a_1+a_2+a_3)f_4f_5}{18(f_1f_5-f_2f_4)f_3}\frac{v_{\varphi}v_{\phi}v_{\Delta}}{\Lambda v_{\chi}}-\frac{(b_1+b_2)f_1f_5-2b_1f_2f_4}{12(f_1f_5-f_2f_4)f_3}\frac{v_{\phi}v_{\Delta}}{\Lambda}\\
\nonumber&&\delta
v_{\varphi2}=\frac{(a_1+a_2+a_3)f_5}{3(f_1f_5-f_2f_4)}\frac{v_{\phi}v_{\Delta}}{\Lambda}-\frac{(b_1-b_2)f_2}{2(f_1f_5-f_2f_4)}\frac{v_{\chi}v_{\phi}v_{\Delta}}{\Lambda
v_{\varphi}}\\
\label{ap32}&&\delta v_{\zeta}=\frac{(a_1+a_2+a_3)f_4}{3(f_1f_5-f_2f_4)}\frac{v_{\phi}v_{\Delta}}{\Lambda}-\frac{(b_1-b_2)f_1}{2(f_1f_5-f_2f_4)}\frac{v_{\chi}v_{\phi}v_{\Delta}}{\Lambda v_{\varphi}}
\end{eqnarray}
In the same way, we obtain the minimization
equations for the shifts $\delta v_{\phi1,2,3}$ and $\delta
v_{\eta1}$
\begin{eqnarray}
\nonumber&&(-2g_1v_{\phi}+g_2v_{\eta})\delta
v_{\phi3}+g_2v_{\phi}\delta
v_{\eta1}+c_1v^2_{\chi}v^2_{\phi}/\Lambda^2=0\\
\nonumber&&(4g_1v_{\phi}+g_2v_{\eta})\delta
v_{\phi2}+c_2v^2_{\chi}v^2_{\phi}/\Lambda^2=0\\
\nonumber&&(-2g_1v_{\phi}+g_2v_{\eta})\delta
v_{\phi1}+c_3v^2_{\chi}v^2_{\phi}/\Lambda^2=0\\
\label{ap33}&&2g_3v_{\phi}\delta v_{\phi3}+2g_4v_{\eta}\delta v_{\eta1}+c_4v^2_{\chi}v^2_{\phi}/\Lambda^2=0
\end{eqnarray}
where the parameters $c_{1,2,3,4}$ are given by
\begin{eqnarray}
\nonumber c_1&=&6s_2+3(s_7+s_8)v_{\eta}/v_{\phi}+2s_{14}v^2_{\varphi}/v^2_{\chi}+(s_{16}+s_{17})v^2_{\varphi}v_{\eta}/(v^2_{\chi}v_{\phi})+(2s_{19}-4s_{20})v_{\varphi}/v_{\chi}\\
\nonumber&&-2(s_{23}+s_{24})v_{\varphi}v_{\eta}/(v_{\chi}v_{\phi})+2s_{28}v_{\varphi}v^2_{\eta}/(v_{\chi}v^2_{\phi})\\
\nonumber
c_2&=&6s_1+3s_6v_{\eta}/v_{\phi}+4s_{13}v^2_{\varphi}/v^2_{\chi}+2s_{15}v^2_{\varphi}v_{\eta}/(v^2_{\chi}v_{\phi})+(2s_{19}-4s_{20})v_{\varphi}/v_{\chi}\\
\nonumber&&-2(s_{23}-s_{24})v_{\varphi}v_{\eta}/(v_{\chi}v_{\phi})+2s_{28}v_{\varphi}v^2_{\eta}/(v_{\chi}v^2_{\phi})\\
\nonumber
c_3&=&6s_2+3(s_7-s_8)v_{\eta}/v_{\phi}+2s_{14}v^2_{\varphi}/v^2_{\chi}+(s_{16}-s_{17})v^2_{\varphi}v_{\eta}/(v^2_{\chi}v_{\phi})+(2s_{19}+8s_{20})v_{\varphi}/v_{\chi}\\
\nonumber&&+4s_{23}v_{\varphi}v_{\eta}/(v_{\chi}v_{\phi})+2s_{28}v_{\varphi}v^2_{\eta}/(v_{\chi}v^2_{\phi})\\
\label{ap34} c_4&=&3r_2+3r_6v^2_{\eta}/v^2_{\phi}+r_8v^2_{\varphi}/v^2_{\chi}+r_{10}v^2_{\varphi}v^2_{\eta}/(v^2_{\chi}v^2_{\phi})+4r_{11}v_{\varphi}/v_{\chi}+2r_{12}v_{\varphi}v_{\eta}/(v_{\chi}v_{\phi})
\end{eqnarray}
Obviously Eq.(\ref{ap33}) admits the solutions
\begin{eqnarray}
\nonumber&&\delta
v_{\phi1}=\frac{c_3}{4g_1}\frac{v^2_{\chi}v_{\phi}}{\Lambda^2}\\
\nonumber&&\delta v_{\phi2}=-\frac{c_2}{2g_1}\frac{v^2_{\chi}v_{\phi}}{\Lambda^2}\\
\nonumber&&\delta v_{\phi3}=\frac{4c_1g_1g_4+c_4g^2_2}{16g^2_1g_4-2g^2_2g_3}\frac{v^2_{\chi}v_{\phi}}{\Lambda^2}\\
\label{ap35}&&\delta v_{\eta1}=\frac{c_1g_2g_3+2c_4g_1g_2}{8g^2_1g_4-g^2_2g_3}\frac{v^2_{\chi}v_{\phi}}{\Lambda^2}
\end{eqnarray}
From the above equations Eq.(\ref{ap35}), we can clearly see that all the shifts $\delta v_{\phi1}/v_{\phi}$, $\delta v_{\phi2}/v_{\phi}$, $\delta v_{\phi3}/v_{\phi}$ and $\delta v_{\eta1}/v_{\eta}$ are of order $\lambda^4_c$.
Finally the equations for the corrections $\delta v_{\Delta1,2,3}$ are
\begin{eqnarray}
\nonumber&&(4h_1v_{\Delta}+h_2v_{\xi})\delta
v_{\Delta1}+d_1v^2_{\chi}v^2_{\Delta}/\Lambda^2=0\\
\nonumber&&(-2h_1v_{\Delta}+h_2v_{\xi})\delta
v_{\Delta3}+d_2v^2_{\chi}v^2_{\Delta}/\Lambda^2=0\\
\label{ap36}&&(-2h_1v_{\Delta}+h_2v_{\xi})\delta
v_{\Delta2}+d_3v^2_{\chi}v^2_{\Delta}/\Lambda^2=0
\end{eqnarray}
where the coefficients $d_{1,2,3}$ are
\begin{eqnarray}
\nonumber
d_1&=&6t_1+3t_6v_{\xi}/v_{\Delta}+(2t_{10}+8t_{12})v_{\varphi}/v_{\chi}+4t_{15}v_{\varphi}v_{\xi}/(v_{\chi}v_{\Delta})+2t_{17}v_{\varphi}v^2_{\xi}/(v_{\chi}v^2_{\Delta})\\
\nonumber&&+4t_{18}v^2_{\varphi}/v^2_{\chi}+2t_{20}v^2_{\varphi}v_{\xi}/(v^2_{\chi}v_{\Delta})\\
\nonumber
d_2&=&6t_2+3t_7v_{\xi}/v_{\Delta}+(2t_{10}-4t_{12})v_{\varphi}/v_{\chi}-2t_{15}v_{\varphi}v_{\xi}/(v_{\chi}v_{\Delta})+2t_{17}v_{\varphi}v^2_{\xi}/(v_{\chi}v^2_{\Delta})\\
\nonumber&&+2t_{19}v^2_{\varphi}/v^2_{\chi}+t_{21}v^2_{\varphi}v_{\xi}/(v^2_{\chi}v_{\Delta})\\
\nonumber
d_3&=&6t_2+3t_7v_{\xi}/v_{\Delta}+(2t_{10}-4t_{12})v_{\varphi}/v_{\chi}-2t_{15}v_{\varphi}v_{\xi}/(v_{\chi}v_{\Delta})+2t_{17}v_{\varphi}v^2_{\xi}/(v_{\chi}v^2_{\Delta})\\
\label{ap37}&&+2t_{19}v^2_{\varphi}/v^2_{\chi}+t_{21}v^2_{\varphi}v_{\xi}/(v^2_{\chi}v_{\Delta})
\end{eqnarray}
The solutions to the above equations Eq.(\ref{ap36}) are given by
\begin{eqnarray}
\nonumber&&\delta v_{\Delta1}=-\frac{d_1}{2h_1}\frac{v^2_{\chi}v_{\Delta}}{\Lambda^2}\\
\nonumber&&\delta
v_{\Delta2}=\frac{d_3}{4h_1}\frac{v^2_{\chi}v_{\Delta}}{\Lambda^2}\\
\label{ap38}&&\delta
v_{\Delta3}=\frac{d_2}{4h_1}\frac{v^2_{\chi}v_{\Delta}}{\Lambda^2}
\end{eqnarray}
It is obvious that $\delta v_{\Delta1,2,3}/v_{\Delta}$ are of order
$\lambda^4_c$, this is because the corrections to the vacuum
alignment of $\Delta$ and $\xi$ arise at the next to next leading
order. In short summary, the modified vacuum configuration of the
flavon fields can be parameterized by Eq.(\ref{ap29}), the shifts $\delta v_{\varphi1}$, $\delta v_{\eta2}$ and
$\delta v_{\xi}$ have been reabsorbed into the redefinition of
$v_{\varphi}$, $v_{\eta}$ and $v_{\xi}$ respectively, which remain
undetermined. The subleading corrections are suppressed by at least
one power of $1/\Lambda$ with respect to the LO results, concretely
$\delta v_{\chi1,2,3}/v_{\chi}$, $\delta
v_{\varphi2}/v_{\varphi}$ and $\delta v_{\zeta}/v_{\chi}$ are of order $\lambda^2_c$, while $\delta v_{\phi1,2,3}/v_{\phi}$, $\delta v_{\eta1}/v_{\eta}$ and $\delta v_{\Delta1,2,3}/v_{\Delta}$ are of
order $\lambda^4_c$. These order of magnitudes can be clearly seen from Eqs.(\ref{ap32},\ref{ap35},\ref{ap38}), we note that the different suppressions of the shifts are due to the constraint of the flavor symmetry $S_4\times Z_3\times Z_4$.
\section*{Appendix C: GUT symmetry breaking}
In the following, we shall briefly discuss the GUT Higgs sector of
the model in the present effective theory. Our Higgs sector is
composed of $H_5$, $H_{45}$, $H_{\overline{5}}$, $H_{\overline{45}}$
and $H_{24}$, the LO $S_4\times Z_3\times Z_4$ invariant
interactions between the different Higgs chiral superfields in the
model are
\begin{eqnarray}
\nonumber&&w_{H}=m_{24}H_{24}H_{24}+\lambda_{24}H_{24}H_{24}H_{24}+\sum^3_{i}f_{Hi}\frac{1}{\Lambda^2}H_{\overline{5}}H_5{\cal O}^{(5)}_i+\sum^3_{i}\lambda_{Hi}\frac{1}{\Lambda^3}H_{\overline{5}}H_{5}H_{24}{\cal O}^{(5)}_i\\
\nonumber&&~~+\sum^2_ic_{Hi}\frac{1}{\Lambda^2}H_{\overline{5}}H_{24}H_{45}{\cal O}^{(6)}_i+b'_{H}\frac{1}{\Lambda^2}H_{\overline{45}}H_{24}H_5\Delta\phi+\sum^3_ib_{Hi}\frac{1}{\Lambda^3}H_{\overline{45}}H_{24}H_{5}{\cal O}^{(7)}_i\\
\label{ap40}&&~~+m_{45}H_{\overline{45}}H_{45}+a_HH_{\overline{45}}H_{45}H_{24}
\end{eqnarray}
where
\begin{eqnarray}
\nonumber&&{\cal O}^{(5)}=\{\Delta^2\chi,\Delta^2\varphi,\Delta\chi\xi\}\\
\nonumber&&{\cal O}^{(6)}=\{\Delta^2,\xi^2\}\\
\label{ap41}&&{\cal O}^{(7)}=\{\chi^3,\chi^2\varphi,\varphi^3\}
\end{eqnarray}
Using the vacuum alignment shown in Eq.(\ref{ap23}) we can immediately obtain that
\begin{eqnarray}
\nonumber&&w_{H}=m_{24}H_{24}H_{24}+\lambda_{24}H_{24}H_{24}H_{24}+f_{H}H_{\overline{5}}H_5+\lambda_{H}H_{\overline{5}}H_{5}H_{24}+c_HH_{\overline{5}}H_{24}H_{45}\\
\label{ap42}&&~~~+b_{H}H_{\overline{45}}H_{24}H_5+m_{45}H_{\overline{45}}H_{45}+a_{H}H_{\overline{45}}H_{45}H_{24}
\end{eqnarray}
with
\begin{eqnarray}
\nonumber&&f_{H}=2f_{H1}\frac{v^2_{\Delta}v_{\chi}}{\Lambda^2}+f_{H3}\frac{v_{\Delta}v_{\chi}v_{\xi}}{\Lambda^2}\\
\nonumber&&\lambda_{H}=2\lambda_{H1}\frac{v^2_{\Delta}v_{\chi}}{\Lambda^3}+\lambda_{H3}\frac{v_{\Delta}v_{\chi}v_{\xi}}{\Lambda^3}\\
\nonumber&&c_{H}=c_{H1}\frac{v^2_{\Delta}}{\Lambda^2}+c_{H2}\frac{v^2_{\xi}}{\Lambda^2}\\
\label{ap43}&&b_{H}=6b_{H2}\frac{v^2_{\chi}v_{\varphi}}{\Lambda^3}+2b_{H3}\frac{v^3_{\varphi}}{\Lambda^3}
\end{eqnarray}
Since all the Higgs fields are neutral under the continuous $U(1)_R$
symmetry, the superpotential Eq.(\ref{ap40}) explicitly break
$U(1)_R$, while preserve the usual R-parity. Certainly we can
construct invariant operators comprising the driving fields, the
Higgs fields and an arbitrary number of flavon fields, however,
these operators don't contribute to the scalar potential due to the
vanishing VEVs of the driving fields. Consequently, To completely
understand the GUT symmetry breaking, maybe we should go beyond the
effective theory framework and consider the ultraviolet completion
\footnote{The same is true for a large class of models with discrete
flavor symmetry, where a continuous $U(1)_R$ symmetry is used to
solve the vacuum alignment problem.}. We note that the effective
superpotential in Eq.(\ref{ap40}) could help us to qualitatively
understand the GUT symmetry breaking, although this approach is not
so satisfactory because of the $U(1)_R$ symmetry breaking. In the
context of the ultraviolet completion of the effective model, the
terms in Eq.(\ref{ap40}) could be generated from a $U(1)_R$
conserving superpotential in which the breaking is mediated by
additional fields which carry $U(1)_R$ charge. The ultraviolet
completion of the model deserves considerable theoretical work
(please see Ref.\cite{Varzielas:2010mp} for an example of the
ultraviolet completion of the $A_4$ model), it is beyond the scope
of the present work.

The scalar potential of the model is determined by the SUSY $F$
terms, $D$ terms and soft terms contributions. We notice that the
first two terms in Eq.(\ref{ap43}) is the interactions for $H_{24}$,
they are exactly the same as those in the conventional GUT theory,
this is because that the Higgs $H_{24}$ is neutral under the flavor
symmetries, consequently the $SU(5)$ GUT symmetry is broken into the
standard model one as usual. Subsequently the VEVs of $H_{5}$,
$H_{45}$, $H_{\overline{5}}$ and $H_{\overline{45}}$ break the
standard model symmetry into the residual $SU(3)_c\times U(1)_{em}$.
Recalling that the parameter $\tan\beta$ could be small or large in
the minimal supersymmetric standard model, this means that a
hierarchy between the Higgs VEVs $v_u$ and $v_d$ can be
accommodated. In exactly the same way, the minor hierarchy between
$v_{\overline{5}}$ and $v_{\overline{45}}$ in Eq.(\ref{19}) can be
achieved by moderately fine-tuning the parameters in the
superpotential $w_H$.


\end{document}